\documentclass[journal]{IEEEtran}
\usepackage{amssymb}
\usepackage{graphicx}
\usepackage{epstopdf}
\usepackage{gensymb}
\usepackage{color}
\usepackage{amsmath}
\usepackage{bm}
\usepackage{etoolbox}
\usepackage{cite}
\usepackage{array}
\usepackage{booktabs}
\usepackage{multirow}
\usepackage{comment}

\usepackage{threeparttable} 
\usepackage{mathtools}
\usepackage{hyperref}
\usepackage{pifont}

\DeclarePairedDelimiter\ceil{\lceil}{\rceil}

\usepackage{url}
\urldef{\mailsa}\path|{alfred.hofmann, ursula.barth, ingrid.haas, frank.holzwarth,|
	\urldef{\mailsb}\path|anna.kramer, leonie.kunz, christine.reiss, nicole.sator,|
	\urldef{\mailsc}\path|erika.siebert-cole, peter.strasser, lncs}@springer.com|    

\usepackage[utf8]{inputenc}
\usepackage[english]{babel}

\usepackage{amsthm}

\theoremstyle{definition}

\definecolor{R}{RGB}{0,0,150}
\definecolor{D}{RGB}{100,0,150}

\theoremstyle{remark}

\ifCLASSINFOpdf
\else
\fi
\begin{document}
\title{Lightweight (Reverse) Fuzzy Extractor with Multiple Referenced PUF Responses}

\author{
	Yansong Gao~{*}, Yang Su{*}, Lei Xu and Damith C.~Ranasinghe     
    \thanks{Corresponding author: Yansong Gao.}
    \thanks{Y.~Gao and Y.~Su contribute equally to this work and share first co-authorship.}
	\thanks{Y.~Gao is with the School of Computer Science and Engineering, NanJing University of Science and Technology (NJUST), China. He is also with Data61, CSIRO, Sydney, Australia. e-mail: yansong.gao@njust.edu.cn.}
    \thanks{L.~Xu is with the the School of Computer Science and Engineering, NanJing University of Science and Technology, China. He is also a visiting scholar at State Key Laboratory of Synthetical Automation for Process Industries. e-mail:xulei\_marcus@126.com.}
	\thanks{Y.~Su, D.~C. Ranasinghe are with the Auto-ID Labs, School of Computer Science, The University of Adelaide, Australia. e-mail: \{yang.su01; damith.ranasinghe\}@adelaide.edu.au.}
}

\maketitle

\begin{abstract}
A Physical unclonable functions (PUF), alike a fingerprint, exploits manufacturing randomness to endow each physical item with a unique identifier. One primary PUF application is the secure derivation of volatile cryptographic keys using a fuzzy extractor comprising of two procedures: i) secure sketch; and ii) entropy extraction. Although the entropy extractor can be lightweight, the overhead of the secure sketch responsible correcting naturally noisy PUF responses is usually costly. We observe that, in general, response unreliability with respect to a enrolled reference measurement increases with increasing differences between the in-the-field PUF operating condition and the operating condition used in evaluating the enrolled reference response. For the first time, we exploit such an \textit{important} but \textit{inadvertent} observation. In contrast to the conventional single reference response enrollment, we propose enrolling \textit{multiple} reference responses (MRR) subject to the same challenge but under multiple distinct operating conditions. The critical observation here is that one of the reference operating conditions is likely to be closer to the operating condition of the field deployed PUF, thus, resulting in minimizing the expected unreliability when compared to the single reference under the nominal condition. Overall, MRR greatly reduces the demand for the expected number of erroneous bits for correction and, subsequently, achieve a significant reduction in the error correction overhead. The significant implementation efficiency gains from the proposed MRR method is demonstrated from software implementations of fuzzy extractors on batteryless resource constraint computational radio frequency identification devices, where realistic PUF data is collected from the embedded intrinsic SRAM PUFs.
\end{abstract}


\IEEEpeerreviewmaketitle
\section{Introduction}\label{Introduction}
Physical unclonable functions (PUFs) exploit manufacturing randomness to create inseparable instance-specific secrets, much like a fingerprint of a human being~\cite{suh2007physical,chang2017retrospective}. The PUF is a promising alternative to low-cost secure key storage. Non-volatile memory (NVM) such as FLASH, predominantly used for digital key storage in electronic components nowadays, may require additional masks and process steps for fabrication. In contrast, silicon PUFs are inherently compatible with standard CMOS fabrication processes, reaping the benefit of reduced manufacturing costs. In addition, a PUF does not store secrets permanently in a digital manner, instead, it utilizes analog randomness to extract secrets on demand. Therefore, the secret is hidden within the physical structure of integrated circuits (ICs) and cannot easily be measured physically; hence, PUF secrets are much less susceptible to invasive attacks in comparison with NVM stored digital secrets~\cite{tuyls2006read,gassend2008controlled,obermaier2018measurement}. This is advantageous when an IC is deployed in a hostile environment where an adversary has physical access to the IC, which is not an unrealistic threat in an Internet of Things (IoT) era. In this new era, PUF based security mechanisms, where a PUF can serve as an inseparable root of trust, are attractive for lightweight authentication and key generation applications to secure low-end IoT devices that lack protected key storage mechanisms.

A PUF is characterized by an instance-specific challenge (incoming binary vector) and response (output binary vector) behavior. The same challenge query applied to different PUF instances produces significantly different responses. Whenever a challenge is repeatedly applied to the same PUF instance, the response should be consistent. However, in reality, consistent response regeneration in not possible since responses are susceptible to noise, such as thermal noise and fluctuation on operating voltage. As a result, the noisy response jeopardizes PUF applications. For key generation, the flipped response bits must be reconciled. As for PUF-based authentication, according to the recent survey of twenty one authentication protocols by Delvaux~\cite{delvaux2017physically}, realizations are classified into two categorizes. The first category falls into the strong PUF obfuscation based authentication, which is a variant of the challenge response pair (CRP) based authentication provided that the relationship between the challenge and response obfuscation holds. Obfuscation is realized through, e.g., randomization, XOR, or decimation~\cite{yu2014noise,rostami2014robust,gao2016obfuscated} but without reliance on a cryptographic primitive, e.g., universal hash function. Unfortunately, it has been demonstrated that it is hard to achieve secure strong PUF obfuscation based authentication, especially in front of various modeling attacks~\cite{ruhrmair2013puf,becker2015pitfalls}. After the recent examination of strong PUF based authentication~\cite{delvaux2017machine}, Delvaux~\cite{delvaux2017machine} indicates that a fairly conservative approach to craft a PUF-based authentication protocol is to convert a noisy response into a stable key and, then follow a keyed algorithm to perform authentication. This approach is classified into the second authentication category~\cite{delvaux2015survey}. We can see that the second approach requires a PUF-based key generator where: i) the response errors are stabilized; and ii) hashed to derive a cryptographic key---together, both procedures are usually termed fuzzy extractor (cf. Section~\ref{sec:RFE}). A fuzzy extractor derives a reliable cryptographic key from noisy raw responses. 

Although, the PUF key generator based authentication assures a high-level of security, the approach is challenged by the high implementation overhead introduced by the the error correction process responsible for stabilizing a noisy response. The prohibitive resource demands for error correction is a significant problem for resource-constraint platforms such as an Internet of Things devices with limited computational capability, memory and power. In this paper, we aim to address this problem. 

We take an important step to investigate novel methodologies to substantially optimize the overhead when implementing a PUF key generator on, especially, resource limited IoT devices (tokens) such as RFID tags and wireless sensors. Our key observation is that all previous PUF key generators solely enroll a single reference response that is evaluated under the so called nominal operating condition, e.g., room temperature. This is ineffective for reducing the unreliability caused by the fact that the operating condition of a PUF in-the-field can vary greatly from the \textit{nominal} operating condition used in the enrollment process\footnote{ We recognize that the study in~\cite{yin2010lisa} conducted Ring Oscillator frequency measurements under two discrete operating conditions with the  objective of maximizing the number of independent response bits enrolled from an ROPUF whilst facilitating the selection of highly reliable bits at a given selection threshold~\cite{yin2010lisa}; however, only the derived single reference is enrolled. }. Conversely, we propose multiple reference response (MRR) enrollment under discrete operating conditions. The crucial observation is that one of the operating conditions of an enrolled reference response will be closer to the operating condition of the PUF in-the-field. Alternatively, though the reproduced response is fixed and is based on the operating condition of the PUF, the reference response can be flexibly selected. The overall result is a significant reduction in the unreliability when compared with the conventional single reference enrollment method.

As an immediate application, we combine MRR with a reverse fuzzy extractor (RFE) to realize a MRR based RFE (MR$^3$FE) that suits lightweight mutual authentication; attributing to the greatly decreased implementation overhead. To examine the MRR method's generalization, it is adopted for a FE, termed  MR$^2$FE. Performance evaluations of both MR$^3$FE and MR$^2$FE are conducted by software implementation on a computational radio frequency identification (CRFID) device that is batteryless and resource limited. For instance, when a key restoration failure rate of less than $10^{-6}$ is desired and pre-selection based MRR enrollment using only three references at \{$-15\celsius$, $25\celsius$, $80\celsius$\} is utilized, MR$^3$FE can reduce the clock cycle overhead by 45\% in comparison with a conventional RFE, while MR$^2$FE can the reduce clock cycle overhead by 42\% in comparison with a conventional FE. 
	
We summarize our main contributions as below:
\begin{itemize}
\item For the first time, we leverage multiple reference response (MRR) enrolled under discrete operating conditions for PUF key generation. As an immediate application, a lightweight mutual authentication protocol based on a reverse fuzzy extractor (RFE), dubbed MR$^3$FE, is proposed. We analyze the key failure rate of MR$^3$FE. 
\item We demonstrate the efficacy of MR$^3$FE to reduce implementation overhead through experiments using software implementations targeting a resource constraint IoT token---a batteryless CRFID device---with an intrinsic SRAM PUF.
\item To examine the generalization of MRR, we experimentally showcase applicability to a fuzzy extractor and also demonstrate the greatly reduced implementation overhead. 
\end{itemize}
\vspace{2mm}

\noindent{\bf Organization:} Section~\ref{sec:background} provides background and related work on FE and RFE, and introduces the conventional RFE-based mutual authentication. Section~\ref{sec:MR3FE} describes MRR enabled RFE-based mutual authentication, which is experimentally validated in Section~\ref{sec:experiment}. Section~\ref{sec:MR2FE} demonstrates the generalization of MRR by adopting it for a FE (MR$^2$FE). We discuss security of MRR when it is employed for a (R)FE in~\ref{sec:discussion}, while we further examine limitations of current investigations and discuss future work. Section~\ref{sec:conclusion} concludes this paper.

\section{Background and Related Work}\label{sec:background}
We begin with a description of the notational format we adopted and give a brief overview of SRAM PUFs. Then we describe related work in the area of fuzzy extractors (FEs) and reverse fuzzy extractors (RFEs) and introduce the conventional RFE-based mutual authentication.

\subsection{Notations}
We denote a vector with a bold lowercase character, e.g., response $\bf r$. We identify an enrolled response from a specific PUF as $\mathbf{r}$, while a reevaluated response from the same PUF is denoted as ${\bf r}^{\prime}$. A matrix is denoted with a bold uppercase character, e.g., a parity check matrix $\bf H$. Functions are printed in sans-serif fonts, e.g., hash function \textsf{Hash()}.

\subsection{SRAM PUF}
There are various silicon PUF constructions that include: delay-based PUF such as Arbiter PUF (APUF)~\cite{gassend2002silicon,zalivaka2016multi} and ring oscillator PUF (ROPUF)~\cite{suh2007physical,cherif2012easy,cao2015low,liu2017acropuf,zhang2014survey}; mismatch based PUFs such as the static random access memory (SRAM) PUF~\cite{holcomb2009power,zeitouniremanence}, latch PUF~\cite{su2008digital}, flip-flop PUF~\cite{maes2008intrinsic,van2010hardware} and Buskeeper PUF~\cite{simons2012buskeeper}; current-based PUF~\cite{majzoobi2011ultra}, and nonlinear current mirror based PUF~\cite{kumar2014design}. Readers are referred to~\cite{gao2015emerging,roel2012physically} for details of various PUF constructions.

This work chooses SRAM PUFs for experimentally demonstrating our MRR methodology. SRAM is pervasively embedded within various electronic commodities. When SRAM is powered up, each SRAM cell has a favored power-up state.
However, the favored power-up state varies from cell to cell, and chip to chip. Therefore, the power-up pattern of SRAM memory can be treated as a PUF where the address of each cell is a challenge and power-up state the response. SRAM PUF is an intrinsic PUF attributing to its wide scale availability and the lack of a requirement for extra hardware overhead~\cite{guajardo2007fpga}; these properties make it one of the most popular silicon PUFs nowadays.

\begin{figure}[t]
	\centering
	\includegraphics[trim=0 0 0 0,clip,width=0.5\textwidth]{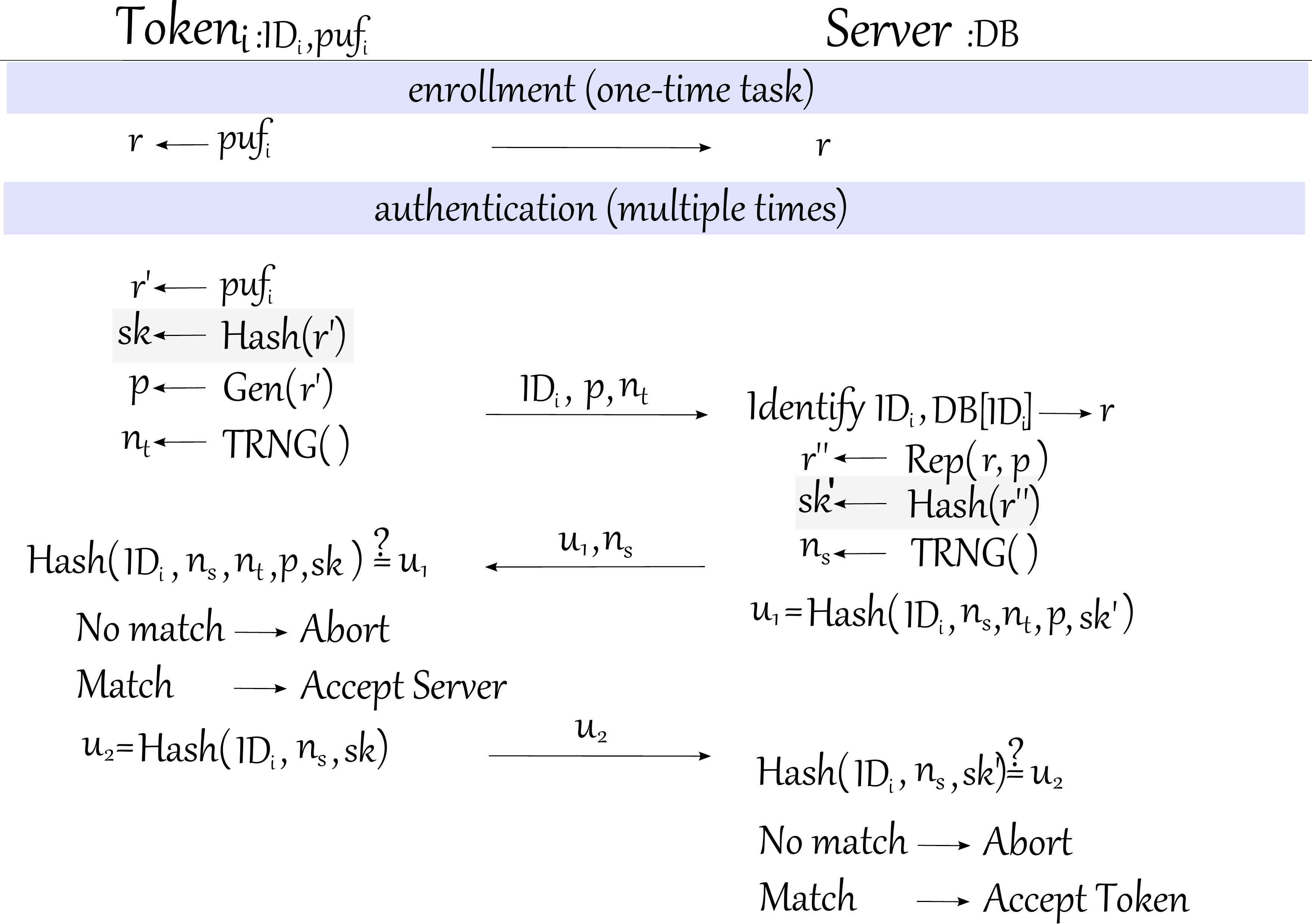}
	\caption{Reverse fuzzy extractor based mutual authentication mechanism.}
	\label{fig:RFE}
\end{figure}

\subsection{(Reverse) Fuzzy Extractor}\label{sec:RFE}
The reproduction of a given PUF response $\bf r$ is not perfect due to its susceptibility to, for example, thermal noise and varying environmental conditions. Thus, raw responses cannot be directly employed as a cryptographic key. A PUF key generator can turn a response $\bf r$ into a cryptographic key with full bit entropy. Usually, a key generator comprises of two procedures: i) secure sketch; and ii) entropy  extraction. Both together are referred to as a fuzzy extractor (FE)~\cite{maes2012pufky,delvaux2015helper,dodis2008fuzzy}. The error correction method deals with generating helper data and the subsequent utility of that data to correct noisy responses. There are two prevalent secure sketch schemes to realize a fuzzy extractor: i) code-offset construction; and ii) syndrome construction~\cite{delvaux2015helper}. We use the syndrome based construction in this paper; we briefly described this construction here.

The secure sketch construction has a pair of functions: \textsf{Gen}() and \textsf{Rep}(). During key enrollment phase, helper data $\bf p$ is computed by using \textsf{Gen}({\bf r}), where ${\bf p}={\bf r}\times {\bf H^T}$ and $\bf H$ is a parity check matrix of a linear error correction code. The key reconstruction described by \textsf{Rep}(${\bf r}^{\prime}$,{\bf p}), where ${\bf r}^{\prime}$ is the reproduced response that may be slightly different from the enrolled response $\bf r$, first constructs a syndrome, ${\bf s}=({\bf r}^{\prime}\times{\bf H^T})\oplus {\bf p}={\bf e}\times{\bf H^T}$, with ${\bf e}$ an error vector. Then through an error location algorithm, $\bf e$ is determined. Subsequently, the response ${\bf r}$ is recovered through ${\bf r}={\bf e}\oplus{\bf r}^{\prime}$. The recovered PUF response $\bf r$ may not ideally be uniformly distributed, therefore, an entropy extraction method such as a universal hash function compresses the PUF response into a cryptographic key with full bit entropy.
	
Normally, in a fuzzy extractor setting, the \textsf{Gen}() function is performed by the server during the provisioning phase to compute helper data. In the field, the \textsf{Rep}() function is implemented on a token. By recognizing that the computational burden of the \textsf{Rep}() function is significantly more than the \textsf{Gen}() function, Van Herrewege {\it et al.}~\cite{van2012reverse} place the \textsf{Gen}() on the resource-constraint token  while leaving the computationally heavy \textsf{Gen}() function execution to the resource-rich server; this method is termed reverse fuzzy extractor (RFE).

\subsection{RFE-based Mutual Authentication}
A reverse fuzzy extractor is beneficial in reducing the implementation overhead of a PUF key generator on a resource limited token. Mutual authentication based on RFE is firstly proposed by Van Herrewege {\it et al.}~\cite{van2012reverse}, later improved by Maes~\cite{roel2012physically}. In Fig.~\ref{fig:RFE}, it depicts the RFE-based mutual authentication protocol in~\cite{roel2012physically}. Notably, the gray shaded secure key ${\bf sk}\leftarrow$\textsf{Hash}($r^\prime$) in Fig.~\ref{fig:RFE} is not explicitly utilized in~\cite{roel2012physically}, instead $r^\prime$ itself is treated as a shared key between the server and the token. Here, instead of using the response $r^\prime$ that might not be uniformly distributed---not having full bit entropy---we adopt the hash function \textsf{Hash()} to extract key {\bf sk} with full bit-entropy.

During the one-time enrollment phase, a response $\bf r$ is enrolled by the server and saved in the database (DB). In the authentication phase, the token computes helper data $\bf p\leftarrow$\textsf{Gen}($\bf r^{\prime}$), where $\bf r^{\prime}$ is the reproduced response. The server receives the public helper data $\bf p$ and uses the enrolled response $\bf r$ to restore the $\bf r^{\prime\prime}\leftarrow$\textsf{Rep}($\bf p$,$\bf r$). Only when the distance between $\bf r^{\prime}$ and $\bf r$ is smaller than a threshold $d$, determined by the error correcting capability of the construction, can $\bf r^{\prime\prime}=$ $\bf r^{\prime}$. Here, only the token and the server share knowledge of the response $\bf r^{\prime}$. Thus, the secret key ${\bf sk}$ given by ${\bf sk}\leftarrow$\textsf{Hash}($\bf r^{\prime}$) is a shared session key. The mutual authentication is realized by employing the nonces $n_t$ and $n_s$ generated by the token's and the server's true random number generators (TRNG), respectively; nonces prevent replaying attacks.
	
Notably, The RFE employed should hold two properties: i) correctness; and ii) security.
\begin{itemize}
\item \textbf{Correctness} implies that the response ${\bf r}^{\prime}$ will be successfully recovered based on the enrolled response $\bf r$ and helper data $\bf p$ through ${\bf r}^{\prime}\leftarrow$\textsf{Rep}($\bf r$,$\bf p$) on the condition that \textsf{FHD}($\bf r$,${\bf r}^{\prime}$)$\le \frac{d}{|\bf r|}$, where \textsf{FHD}() evaluates fractional Hamming distance (FHD) between two binary vectors.
\item \textbf{Security} implies that given the exposed helper data $p$, there is adequate residual entropy in the generated response $r^{\prime}$. 
\end{itemize} 
	
Our focus is on the correctness requirement as we are aiming to significantly reduce the \textsf{Gen}() function implementation overhead on a token based on the MRR method. Although our work focuses on the application of MRR to present the multiple referenced response based reverse fuzzy extractor (MR$^3$FE), our work is not intending to invent any methodology to enhance the security of the RFE-based mutual authentication mechanism, we simply inherit its security~\cite{maes2016secure,helperdatamultiple2017,eliminate2017}. Nonetheless, for completeness, we discuss the security of (reverse) fuzzy extractors in Section~\ref{sec:discussion}.
	
\section{Multiple Referenced Response Based Reverse Fuzzy Extractor (MR$^3$FE)}\label{sec:MR3FE}
In this sections we explain our intuition for developing the multiple reference response (MRR) approach, in general, and then focus on the application of the approach in its most interesting context, a reverse fuzzy extractor (RFE). We explain our rationale by developing an understanding of response unreliability.

The commonly used PUF reliability model, e.g., in~\cite{maes2012pufky,yulockdown}, assumes a fixed error rate, specifically, each response reevaluation is assigned with the same error rate. This is also referred to as homogeneous response error rate. In practice, PUF responses are experimentally demonstrated to exhibit a bit-specific reliability---heterogeneous error rate~\cite{maes2013accurate,maes2009soft}. In this study, we consider the expected value of BER as in~\cite{yulockdown} since this provides a convenient but valid method to analyze the key failure rate in relation to a (reverse) fuzzy extractor. Now, we can express BER as:
\begin{equation}\label{eq:BER}
{\rm BER}=\textsf{E}(\textsf{FHD}({\bf r},{\bf r}^{\prime})),
\end{equation} 
where $\bf r$ and ${\bf r}^{\prime}$ are two distinct and random response evaluations subject to the same challenge applied to the same PUF. Here \textsf{E}() is the expectation operator. 
Commonly, $\bf r$ is a reference response evaluated under a given operating condition and ${\bf r}^{\prime}$ is the reproduced response evaluated, most likely under a differing operating condition. BER is influenced by factors such as thermal noise as well as environmental parameters e.g., supply voltage and temperature. 
\begin{figure}[h]
	\centering
	\includegraphics[trim=1.8cm 0 0 0,clip,width=0.52\textwidth]{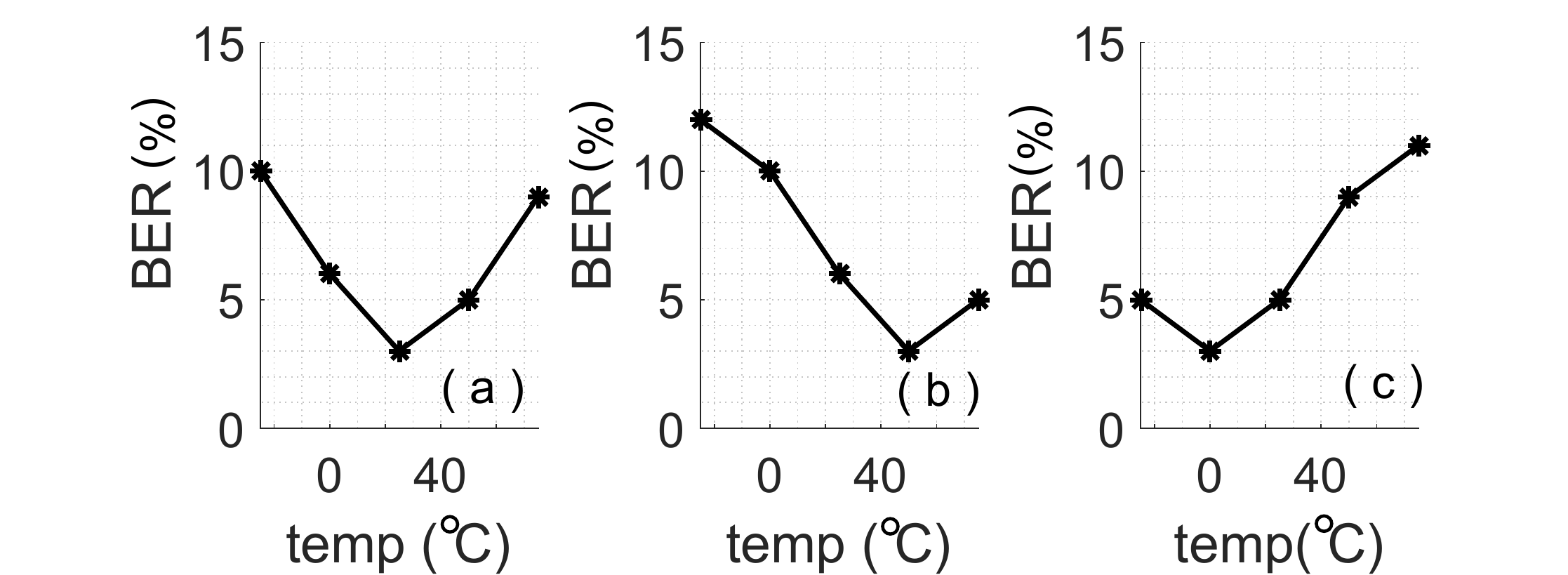}
	\caption{(a) The reference response is enrolled under a nominal operating condition of $25\celsius$. To the best of our knowledge, all current PUF applications enroll only a single response, e.g., evaluated under room temperature. (b) Reference response is enrolled under $50\celsius$. (c) Reference response is enrolled under $0\celsius$.}
	\label{fig:BERtwoOC}
\end{figure}

We use an example to explain our observations and rational. Fig.~\ref{fig:BERtwoOC}(a)\footnote{The BER value in this figure is not obtained from experimental evaluations, it is only for illustrative purpose.} illustrates a {\it single} reference response enrolled under $25\celsius$\footnote{Supply voltage is constant.} that is the nominal reference operating condition. We can see that the BER increases when the operating temperature deviates away from the reference operating condition of $25\celsius$. The maximum BER is around 10\%, which occurs at $-25\celsius$. The minimum BER is under the reference temperature of $25\celsius$. This minimum BER is solely caused by thermal noise. In Fig.~\ref{fig:BERtwoOC}(b), the {\it reference response} is enrolled under $50\celsius$. We can see that the minimum BER appears at the $50\celsius$; the nominal reference operating condition in this case. The maximum BER is approximately 12\% when the regenerated response is evaluated under $-25\celsius$ that is $75\celsius$ below the reference operating condition. Similarly, In Fig.~\ref{fig:BERtwoOC}(c), when the reference response is enrolled under $0\celsius$, the minimum BER occurs at $0\celsius$ and the maximum BER around 12\% occurs when the operating condition increases by $75\celsius$.

In summary, no matter which specific nominal reference operating condition is selected, for example, $-25\celsius$, $25\celsius$ or $50\celsius$, the minimum BER is always achieved at the reference nominal operating condition. BER increases as the difference between the reference nominal operating condition and the operating condition under which response ${\bf r}^{\prime}$ is reproduced increases. One important fact we observe is that BER is highly related to the selection of the reference operating condition and the operating condition of the PUF in the field. A deviation of the operating conditions of the PUF in the field from that under which a response is enrolled will always lead to a deterioration in the expected BER.  Although we cannot change the operating condition under which the PUF operates in the field, we recognize that we can potentially choose a suitable {\it reference} operating condition during response reconciliation to reduce the maximum number of erroneousness bits we expect in a regenerated response. Next, we utilize this important observation to reason the multiple reference response based RFE mutual authentication (MR$^3$FE) mechanism.

\subsection{RFE based Mutual Authentication with MRR}
\begin{figure}[h]
	\centering
    \includegraphics[trim=0 0 0 0,clip,width=0.5\textwidth]{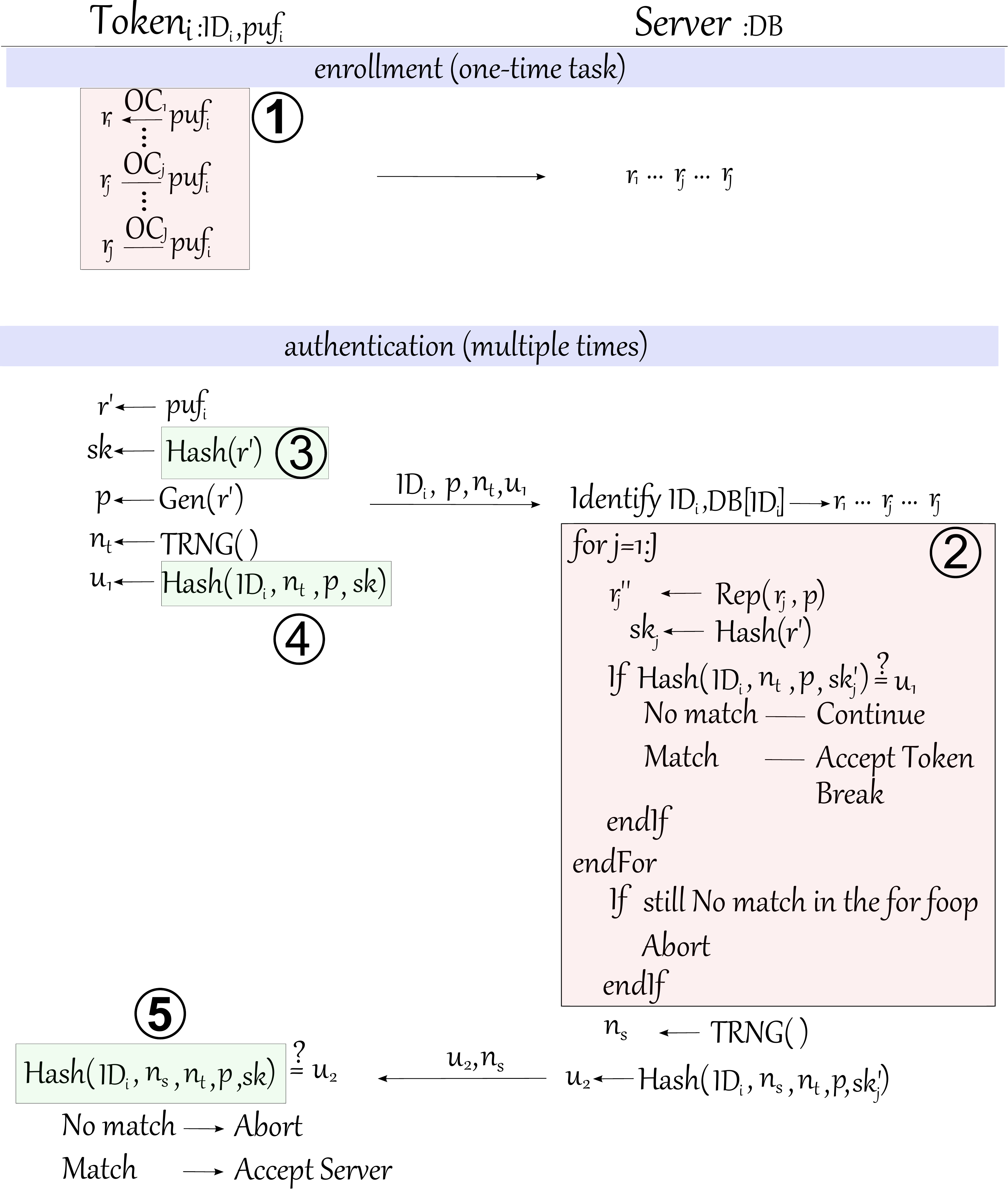}
	\caption{MRR based RFE mutual authentication. OC stands for operating condition.}
	\label{fig:MRRFE}
\end{figure}
Fig.~\ref{fig:MRRFE} depicts the proposed MR$^3$FE mutual authentication protocol. In comparison with conventional RFE based mutual authentication (cf. Fig~\ref{fig:RFE}), there are two distinction differences:
\begin{itemize}
	\item In the enrollment phase, instead of enrolling a {\it single} reference response, the server enrolls multiple reference responses; each reference response is evaluated at a different operating condition. This is highlighted in $\text{\ding{172}}$.
	\item In the authentication phase, the server recovers the regenerated response ${\bf r}^{\prime}$ of the token based on the enrolled multiple reference responses. This is highlighted in $\text{\ding{173}}$.
\end{itemize}

Next we elaborate on the MR$^3$FE mutual authentication by taking two reference responses as an example.
\vspace{3mm}

\noindent{\bf An Example with Two Reference Responses:~} In Fig.~\ref{fig:BERtwoOC}, during the enrollment phase, we assume that the server enrolls two reference responses, ${\bf r}_1$ and ${\bf r}_2$, evaluated under $50\celsius$ and $0\celsius$, respectively. It is worth reminding that ${\bf r}_1$ and ${\bf r}_2$ are subject to the same challenge applied to the same PUF.

In the authentication phase, the token reproduces the response ${\bf r}^{\prime}$ and then computes the corresponding helper data ${\bf p}\leftarrow$\textsf{Gen}(${\bf r}^{\prime}$). In addition, verification data ${\bf u}_1 \leftarrow$\textsf{Hash}(ID$_i$, $n_t$, $\bf sk$, $\bf p$) is computed, where ${\bf u}_1$ is a keyed hash value with $\bf sk$ as the key. The ID$_i$ is the ID of current token, $n_t$ is a nonce generated by the token. ID$_i$, $n_t$, $\bf p$ along with the ${\bf u}_1$ are publicly sent to the server.
	
The server now attempts to reconstruct the response ${\bf r}^{\prime}$ based on its enrolled responses: ${\bf r}_1$ and ${\bf r}_2$. This can be handled in an iterative way. To be precise, the server first uses ${\bf r}_1$ to generate ${\bf r}^{\prime \prime}\leftarrow$\textsf{Rep}(${\bf r}_1$, $\bf p$). Once  response ${\bf r}^{\prime \prime}$ is obtained, the server verifies whether \textsf{Hash}(ID$_i$, $n_t$, ${\bf sk}^\prime$, $\bf p$) equals ${\bf u}_1$ with secret key ${\bf sk}^{\prime} \leftarrow$\textsf{Hash}(${\bf r}^{\prime \prime}$) and ${\bf u}_1$ the verification value sent by the token. If the verification is successful, then ${\bf sk}={\bf sk}^{\prime}$, for this reason, the $ {\bf r}^{\prime} $ is deemed to be successfully restored. Mutual authentication can now proceed based on the shared secret session key $ \bf sk $. If \textsf{Hash}(ID$_i$, $n_t$, ${\bf sk}^{\prime}$, $\bf p$) is not equal to ${\bf u}_1$ and the verification fails, the server continues to use ${\bf r}_2$ for reconstructing $ {\bf r}^{\prime} $ to determine whether $\bf r_{\prime}$ can be successfully recovered. 
	
Notably, it is only after {\it both} ${\bf r}_1$ and ${\bf r}_2$ are exhausted in the recovery of the response $ {\bf r}^{\prime}$ that MR$^3$FE based mutual authentication fails. This occurs on the condition that the verification of ${\bf u}_1$ has \textit{failed} and implies that the recovery of $ {\bf r}^{\prime} $ has failed.
\vspace{3mm}

\noindent\textbf{Advantages:~} Following the two reference response example above, advantages of MR$^3$FE are clear. Let us first assume that the computed helper data $\bf p$ by the token is only able to guarantee a successful secret key $\bf sk$ computation by the server with $ {\bf sk} \leftarrow$\textsf{Hash}($ {\bf r}^{\prime} $) {\it when the BER is no more than 5\%}---in other words, less than 5\% of response bits display errors under reevaluation. Assume a {\it single} reference response $\bf r$ under $25\celsius$ is utilized for key reconstruction as in the conventional RFE case, but response $ {\bf r}^{\prime} $ is reproduced under $-25\celsius$. We can observe from Fig.~\ref{fig:BERtwoOC}(a) that the $ {\bf r}^{\prime} $ is highly unlikely to be correctly recovered by the server because the BER for the reference response $\bf r$ evaluated under $25\celsius$ is much higher than 5\% at  $-25\celsius$.

Let's now assume employing the two reference responses, ${\bf r}_1$ and ${\bf r}_2$, as in MR$^3$FE, and still assume that $ {\bf r}^{\prime} $ is from the PUF operating under $-25\celsius$. We can see that reference response ${\bf r}_2$ has a high chance to {\it successfully} recover response $ {\bf r}^{\prime} $ relying on the fact that the BER using ${\bf r}_2$ evaluated under $0\celsius$ as a reference response is now less than 5\%---see Fig.~\ref{fig:BERtwoOC}(c). Similarly, if $ {\bf r}^{\prime} $ is from $75\celsius$, then using ${\bf r}_1$ evaluated under $50\celsius$ as a reference response will lead to a BER of less than 5\%, see Fig.~\ref{fig:BERtwoOC}(c) and consequently to a successful response recovery. 
\vspace{2mm}

{\it Overall, we can observe by using MRR, though the server is unable to change the operating condition under which the regenerated ${\bf r}^{\prime}$ is evaluated, the server possesses the capability to employ an appropriate reference response to minimize the expected difference between a reference response $\bf r$ and the regenerated response ${\bf r}^{\prime}$ to meet a given error correcting capability threshold $d$.}
\vspace{2mm}

Next, we analyze the key reconstruction failure rate of MR$^3$FE mutual authentication, which is also the false rejection rate of the authentication mechanism.

\subsection{Key Failure Rate}\label{sec:failRate}
To validate the efficiency of the proposed MR$^2$FE and MR$^3$FE, we focus on the {\it average} failure rate of the PUF key generator. In~\cite{maes2013accurate}, it is demonstrated that the expected value of key failure rate based on a bit specific reliability model is equivalent to the key failure rate predicted under the commonly used reliability model with a fixed response error rate model. In other words, the homogeneous reliability model does correctly capture the average key failure rate of a PUF key generator~\cite{maes2013accurate,delvaux2015helper}. Therefore we will use BER defined in \eqref{eq:BER} to express key failure rate.

Our study uses the family of BCH($n$, $k$, $t$) linear codes with a syndrome based decoding strategy to realize a reverse fuzzy extractor considering its popularity~\cite{maes2012pufky,delvaux2015helper} and its security property~\cite{delvaux2015helper,becker2017robust}---we discuss security of fuzzy extractors in Section~\ref{sec:discussion}. Here, $n$ is the codeword length, $k$ is the code size, $t$ is the number of errors that can be corrected within this $n$-bit block. 
Assuming response bit errors are independently and identically distributed (i.i.d.), we can express the average key failure rate of recovering an $n$-bit  response ${\bf r}^{\prime} $ based on a selected reference response ${\bf r}_j$, termed as $\mathbb{P}_{1j}$, where the $j\in \{1,..,J\}$ with $J$ as the number of multiple references employed by the server, as:
\begin{equation}
\mathbb{P}_{1j}= 1 - \textsf{F}_B(t;n,{\rm BER}_j)
\end{equation}
where BER$_j$ is the BER using ${\bf r}_j$ as the reference response. Here, \textsf{F}$_B()$ is a cumulative density function of a binomial distribution with $t$ successes in $n$ Bernoulli trials, with each trial having success probability of $p$, expressed as:
\begin{equation}
\textsf{F}_B(t;n,p)=\sum^{t}_{t=0}{n \choose t}p^t(1-p)^{(n-t)}.
\end{equation}

A BCH($n$, $k$, $t$) encoding produces ($n-k$)-bit helper data assumed to be publicly known while $k$ bits form the secret key material. For a single BCH($n$, $k$, $t$) block, the complexity of finding the $k$-bit response from ${\bf r}^{\prime} $ is $2^{k}$. It is not common to use a single large BCH($n$, $k$, $t$) block; typically a large block is split into small processing blocks to reduce implementation complexity~\cite{hiller2016cherry}. For $k$ bits of key material, response $ {\bf r}^{\prime} $ can be divided into multiple non-overlapping blocks of a BCH($n_1$, $k_1$, $t_1$) code where $n_1<n$ and $k_1<k$ for a parallel implementation. Now the complexity of finding the $k$ bit secret is $2^{k_1\cdot L}$ where $L$ is the number of parallel BCH($n_1$, $k_1$, $t_1$) code blocks used to realize $k$ bits of secret key material. Given a BCH($n_1$, $k_1$, $t_1$) code employed to gain a security level of $k$ bits with $L=\ceil*{k/k_1}$ blocks, the key recovery failure rate under the assumption of i.i.d code blocks is:
\begin{equation}\label{eq:Pfail-Lblocks}
\mathbb{P}_{2j}= 1 - (1-\mathbb{P}_{1j})^L.
\end{equation}

When all $J$ reference responses $\{ {\bf r}_1,...,{\bf r}_j,...,{\bf r}_J \}$ are employed, ${\bf r}^{\prime}$ reconstruction fails only when {\it all} reference responses cannot restore the response $ {\bf r}^{\prime} $. Therefore, the key failure rate ${\mathbb{P}_{\rm fail}}$ for $J$ reference responses can be expressed as a joint probability distribution:
\begin{equation}
{\mathbb{P}_{\rm fail}}=Pr({\bf r}_1,\cap...\cap,{\bf r}_j,\cap...\cap,{\bf r}_J).
\end{equation}

However, due to the complexity of PUF response properties, e.g., correlations, formally deriving a joint distribution without assuming that $\{ {\bf r}_1,...,{\bf r}_j,...,{\bf r}_J \}$ are independently drawn under distinct operating conditions is a non-trivial task\footnote{Under an assumption of independence, the key failure rate ${\mathbb{P}_{\rm fail}}= \prod_{j=1}^{J} \mathbb{P}_{\rm 2_j}$.}. We propose using a very conservative evaluation of the key failure rate ${\mathbb{P}_{\rm fail}}$ without a prior notion of independent implied on the reference responses $\{ {\bf r}_1,...,{\bf r}_j,...,{\bf r}_J \}$. We recognize that we can express the upper bound of the key failure rate $\mathbb{P}_{\rm fail}$ as:
\begin{equation}
{\mathbb{P}_{\rm fail}}=Pr({\bf r}_1,\cap...\cap,{\bf r}_j,\cap...\cap,{\bf r}_J) \leq min\{\mathbb{P}_{2j}\},~j\in\{1,...,J\}
\end{equation}
Now we adopt the very conservative estimate: 
\begin{equation}
{\mathbb{P}_{\rm fail}}= min\{\mathbb{P}_{2j}\},~j\in\{1,...,J\}
\end{equation} 
in our analysis.

\section{Experimental Validations}\label{sec:experiment}	
\begin{figure}
	\centering
	\includegraphics[trim=0 0 0 0,clip,width=0.50\textwidth]{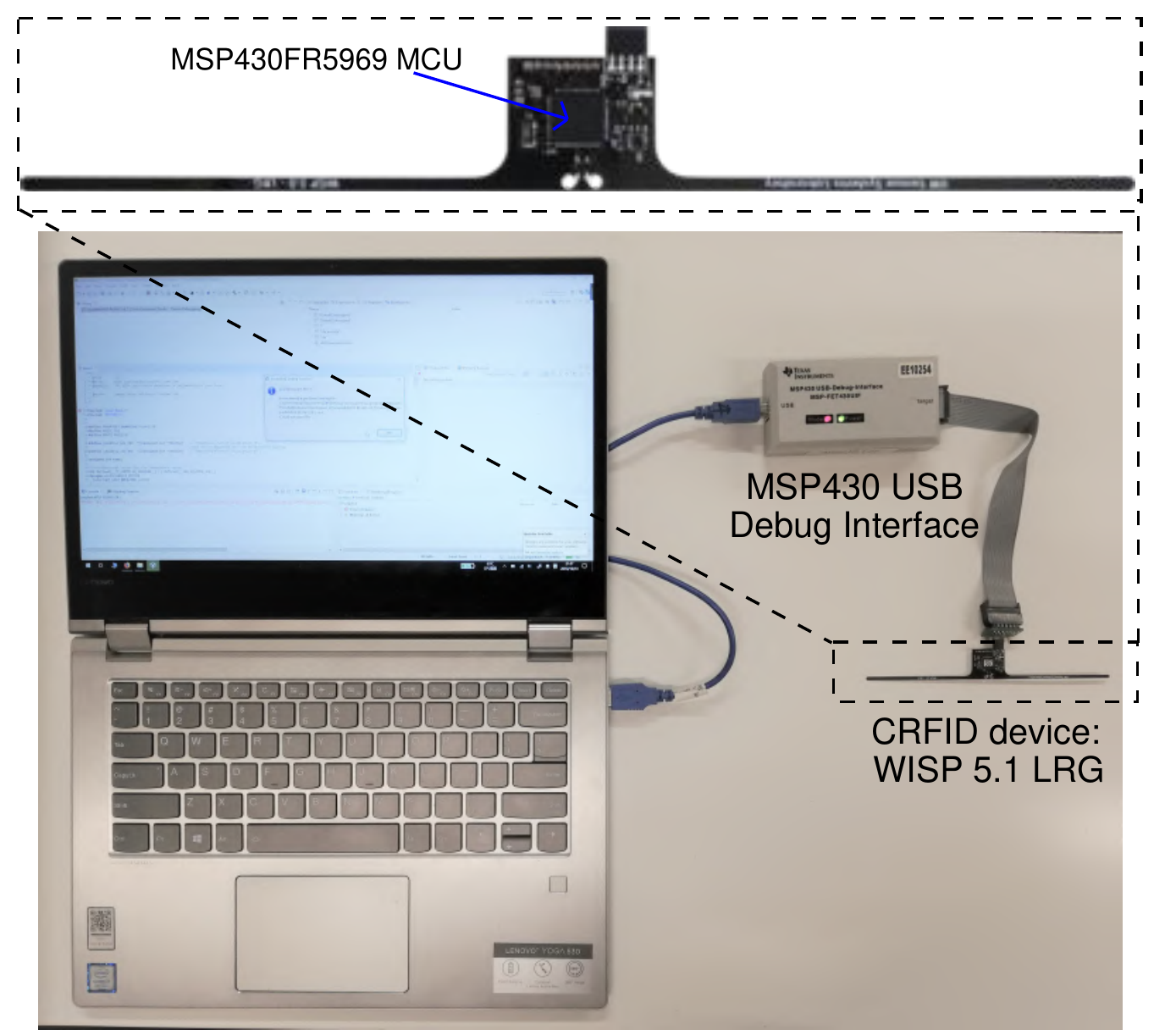}
	\caption{A laptop running Code Composer Studio (CCS) and connected to a USB based JTAG interface for debugging and programming the CRFID device.}
	\label{fig:SetupPic}
\end{figure}

We employ the ultra low power microcontrollers used in  CRFID transponders (WISP5.1LGR) to evaluate the overhead of the proposed MR$^3$FE mutual authentication mechanism as illustrated in Fig.~\ref{fig:SetupPic}. The battery-less CRFID transponder is a highly resource constrained device that operated under harvested power from radio frequency energy. A CRFID device is representative of a low-end resource limited IoT device. Since a CRFID device has SRAM memory, it has the potential to use an intrinsic SRAM PUF as a trust anchor without requiring additional hardware~\cite{Yang2018scode}. 
\subsection{SRAM PUF Dataset}
The PUF CRP dataset used is from 23 MSP430FR5969 microcontrollers (MCUs) used in CRFID transponders (cf. Fig.~\ref{fig:SetupPic}). From each MCU, we read power-up states of 16,384 (2KB) SRAM cells as SRAM PUF responses. It has been experimentally shown that the SRAM PUF reliability is much less sensitive to voltage variations compared with temperature fluctuations attributing to the SRAM cell's symmetric structure~\cite{selimis2011evaluation,roel2012physically,xiao2014bit}. Hence, we focus on its reliability under varying temperature conditions: $-15\celsius$, $0\celsius$, $25\celsius$, $40\celsius$ and $80\celsius$. Under each temperature condition, each response bit is repeatedly measured 100 times.

\subsection{Overhead Evaluations}
\noindent\textbf{Test Setup:~}\label{sec:TestSetup}
The test environment used is Texas Instruments' (TI) Code Composer Studio 7.2.0, the C code used is downloaded to a MSP430FR5969 LaunchPad Evaluation Kit via USB. TI CCS has a built-in GCC toolchain for our hardware kit. This includes the \texttt{msp430-gcc-6.4.0.32} win32 compiler. Considering that our main purpose is to demonstrate enhanced efficacy of MR$^3$FE compared to the conventional RFE in a {\it relative} manner, dedicated optimization of the C code was deemed out of scope. We agree that  optimization~\cite{aysu2016compact} of the fuzzy extractor code can be carried out to further minimize the {\it absolute} implementation overhead of the MR$^3$FE.

The software instructions are executed sequentially as advanced out-of-order execution is unavailable for typical resource-constraint MCUs. The overhead measured in terms of clock cycles to complete the algorithm is our primary concern. We measured clock cycles using Profile Clock tool supported in the CCS environment. In addition, we also measure memory usage. Besides the 2~KB SRAM memory embedded in the MSP430FR5969 microcontroller, it is configured with a 63~KB Ferroelectric Random Access Memory (FRAM). Here {\bf FRAM usage} (overhead) is reflective of code size, while the {\bf SRAM usage} represents size of the internal state used by the algorithm. The code size is assessed by the \texttt{.text} block in FRAM using Memory Allocation tool in CCS, the internal state is manually counted for any local variable declared inside the algorithm routine.

Hash function and BCH code encoding are two pivotal components for realizing the MR$^3$FE and dominates overhead of MR$^3$FE implementation. We comprehensively evaluate these building blocks by testing:
\begin{itemize}
\item \textbf{Hash Functions:~} Six different hash functions are tested. The results are listed in Table~\ref{tab:hash} in the Appendix. We evaluate clock cycles and memory overhead. The input message size we selected is 240 bytes for these tests.
Among all six software based hash implementations, the BLAKE2s-128 showed the best performance with a 128-bit hash. Therefore we selected BLAKE2s-128 for our evaluations.
\item \textbf{BCH Code Encoding:~} BCH($ n_1 $, $ k_1 $, $ t_1 $) code encoding overhead under different $ n_1 $, $ k_1 $, $ t_1 $ settings are tested. Results are detailed in Table.~\ref{tab:BCH} in the Appendix.
\end{itemize}

\subsection{Comparisons}\label{Sec:M3RFECompare}
\noindent\textbf{BER:~}
We first evaluate BER under three different response enrollment approaches: i) single readout; ii) majority voting; and iii) pre-selection.
\begin{itemize}
\item In the single readout response enrollment, all the enrolled responses under a distinct temperature is evaluated only \textit{once}.
\item In the majority voting response enrollment, all the responses under a distinct temperature are evaluated 9 times and then the majority vote is applied for enrollment.
\item In the pre-selection response enrollment, first, each response under $25\celsius$ is repeatedly measured 10 times, only the response bits exhibiting 100\% reliable regenerations (all `1's/`0's) are selected---we discarded 12\% of bits during this process. Then the reference responses under other temperatures, ${ -15\celsius, 0\celsius, 40\celsius, 80\celsius } $ are obtained by applying majority voting to the preselected responses using 9 repeated measurements.
\end{itemize}

\begin{figure}[h!]
	\centering
	\includegraphics[trim=2.0cm 0 2.0cm 0,clip,width=0.5\textwidth]{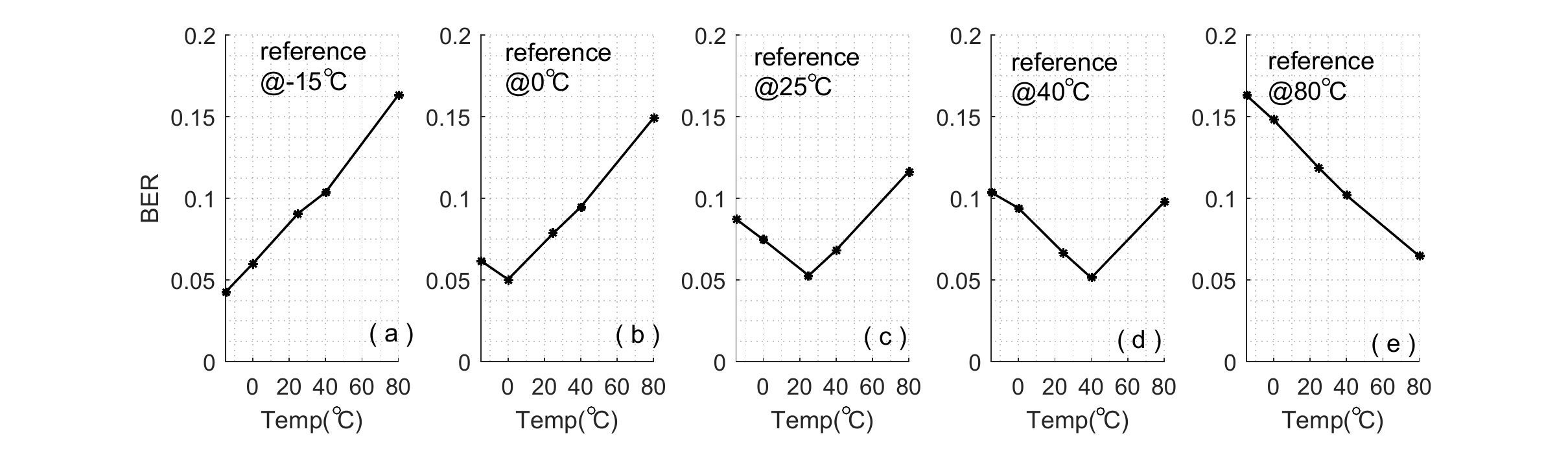}
	\caption{BER when single readout response enrollment is utilized. (a) Reference response is enrolled at $-15\celsius$. (b) Reference response is enrolled at $0\celsius$. (c) Reference response is enrolled at $25\celsius$. (d) Reference response is enrolled at $40\celsius$. (e) Reference response is enrolled at $80\celsius$.}
	\label{fig:BER5OC}
\end{figure}

\begin{figure}[h!]
	\centering
	\includegraphics[trim=1.5cm 0 1.5cm 0,clip,width=0.5\textwidth]{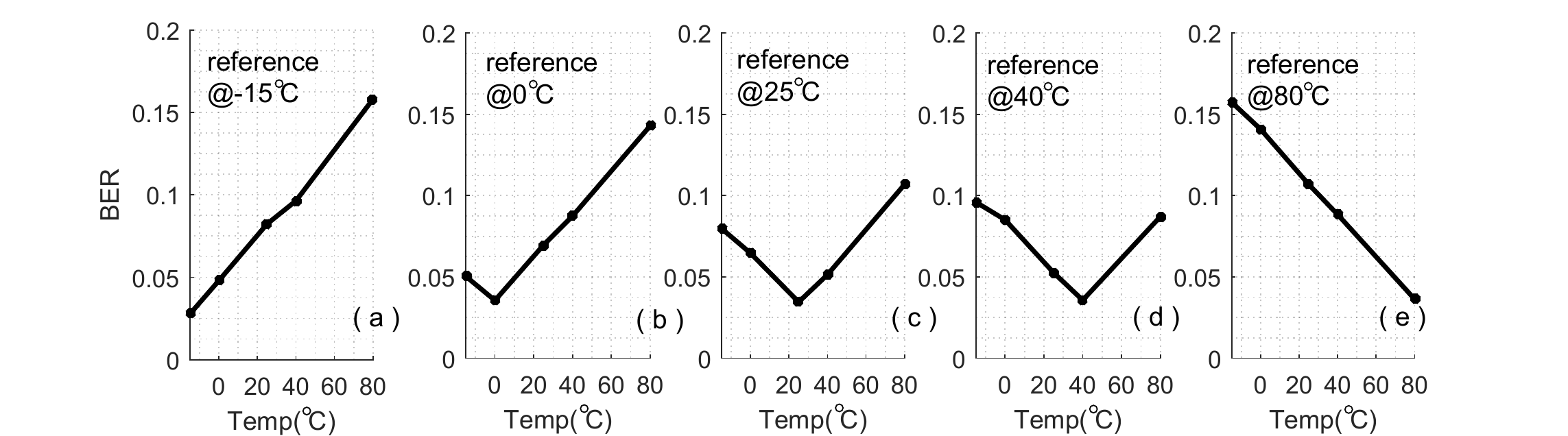}
	\caption{BER when majority voting response enrollment is utilized. (a) Reference response is enrolled at $-15\celsius$. (b) Reference response is enrolled at $0\celsius$. (c) Reference response is enrolled at $25\celsius$. (d) Reference response is enrolled at $40\celsius$. (e) Reference response is enrolled at $80\celsius$.}
	\label{fig:BER5OC_V1}
\end{figure}

\begin{figure} [h!]
	\centering
	\includegraphics[trim=0cm 0 1cm 0,clip,width=0.45\textwidth]{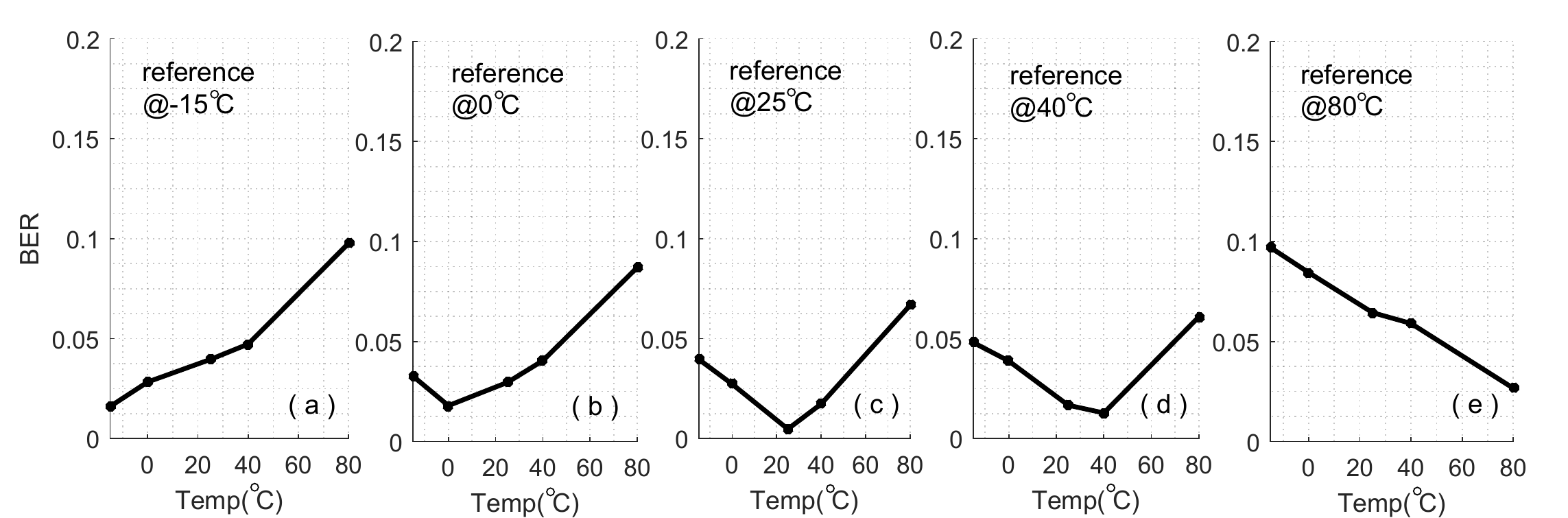}
	\caption{BER when preselection response enrollment is utilized. (a) Reference response is enrolled at $-15\celsius$. (b) Reference response is enrolled at $0\celsius$. (c) Reference response is enrolled at $25\celsius$. (d) Reference response is enrolled at $40\celsius$. (e) Reference response is enrolled at $80\celsius$.}
	\label{fig:BER5OC_preselection}
\end{figure}

BER evaluations based on the three different response enrollment approaches we employed---single readout, majority voting and pre-selection---are illustrated in Fig.~\ref{fig:BER5OC}, Fig.~\ref{fig:BER5OC_V1} and Fig.~\ref{fig:BER5OC_preselection}, respectively.

We observe the following:
\begin{itemize}
\item Regardless of response enrollment approach, it is empirically verified that the BER increases as a function of the temperature difference between the response regeneration temperature and the reference temperature. 
\item As expected, both majority voting and pre-selection approaches reduce BER; the pre-selection approach being the most effective.
\end{itemize}
\vspace{2mm}

\noindent\textbf{Key Failure Rate:~} Based on BER values obtained from the three different response enrollment approaches, we are able to evaluate the key failure rate. We used Parallel BCH($ n_1 $, $ k_1 $, $ t_1 $) blocks as discussed in Section~\ref{sec:failRate}. We consider an evaluation under the assumption of deriving a 128 bit secret. Therefore, we determine the number of BCH($n_1, k_1, t_1$) blocks required by using $\left \lceil \frac{128}{k_1} \right \rceil$. The key failure rates we have determined is detailed in Table.~\ref{tab:FEKeyGen}. We observe the following: 
\begin{itemize}
\item Before applying MRR, majority voting and pre-selection reduces the BER and thus decreases the key failure rate.
\item Regardless of response enrollment approaches, our MRR approach further suppresses the key failure rate. In other words, the MRR approach complements response reliability enhancement approaches such as majority voting and pre-selection performed in the \textit{enrollment phase}.
\end{itemize}
\vspace{2mm}

\begin{table*}[h]
	\centering 
	\caption{Key failure rate achieved for single readout, majority voting and preselection response enrollment approaches to realize a 128-bit key.}
	\resizebox{1.0\textwidth}{!}{
		\begin{tabular}{c | c | c | c | c | || c | c | c || c | c | c }
			\toprule 
			\toprule 
			& & \multicolumn{3}{c||}{ Single readout} & \multicolumn{3}{c||}{ Majority voting} & \multicolumn{3}{c}{ Preselection} \\
			\cmidrule(l){3-5} 
            \cmidrule(l){6-8}
            \cmidrule(l){9-11}
            
			($ n_1 $,$ k_1 $,$ t_1 $) & \rotatebox{0}{block num.}  & SRR & 2MRR & 3MRR & SRR & 2MRR & 3MRR & SRR & 2MRR & 3MRR \\ 
            \midrule
			(63,18,10)	& 8 & 0.6074 & 0.2821 & $2.67\times 10^{-2}$ & 0.4355 & 0.1446 & $2.7\times 10^{-3}$ & $2.26\times 10^{-2}$ & $1.10\times 10^{-2}$ & $8.21 \times 10^{-6}$ \\ 
			\midrule
			(63,16,11)	& 8 & 0.3789 & 0.1342 & $8.2\times 10^{-3}$ & 0.2366 & $5.9\times 10^{-2}$ & $6.22\times 10^{-4}$ & $6.8\times 10^{-3}$ & $3.0\times 10^{-3}$ & $9.85 \times 10^{-7}$ \\ 
			\midrule
			(127,29,21)	& 5 & 0.1712 & $2.86\times 10^{-2}$ & $2.49\times 10^{-4}$ & $7.55\times 10^{-2}$ & $7.1\times 10^{-3}$ & $2.82\times 10^{-6}$ & $1.79\times 10^{-4}$ & $4.36\times 10^{-5}$ & $2.97\times 10^{-11}$\\ 
			\midrule
			(127,22,23)	& 6 & $6.62\times 10^{-2}$ & $7.4\times 10^{-3}$ & $3.04\times 10^{-5}$ & $2.39\times 10^{-2}$ & $1.4\times 10^{-3}$ & $1.91\times 10^{-7}$ & $2.08\times 10^{-5}$ & $4.19\times 10^{-6}$ & $5.47\times 10^{-13}$\\ 
			\midrule
			(127,15,27)	& 9 & $5.7\times 10^{-3}$ & $2.95\times 10^{-4}$ & $2.66\times 10^{-7}$ & $1.4\times 10^{-3}$ & $3.51\times 10^{-5}$ & $5.09\times 10^{-10}$ & $1.66\times 10^{-7}$ & $2.67\times 10^{-8}$ & $< 10^{-21}$ \\ 
			\midrule
			(255,47,42)	& 3 & $2.48\times 10^{-2}$ & $9.0\times 10^{-4}$ & $1.25\times 10^{-7}$ & $5.4\times 10^{-3}$ & $6.8\times 10^{-5}$ & $2.47\times 10^{-11}$ & $6.69\times 10^{-8}$ & $4.58\times 10^{-9}$ & $< 10^{-21}$ \\ 
			\midrule
			(255,29,47)	& 5 & $2.8\times 10^{-3}$ & $3.96\times 10^{-5}$ & $8.27 \times 10^{-10}$ & $3.83\times 10^{-4}$ & $1.62\times 10^{-6}$ & $3.66 \times 10^{-14}$ & $3.92\times 10^{-10}$ & $1.65\times 10^{-11}$ & $< 10^{-21}$ \\ 
			\midrule
			(255,21,55)	& 7 & $1.52\times 10^{-5}$ & $4.72\times 10^{-8}$ & $4.59 \times 10^{-14}$ & $9.90\times 10^{-7}$ & $7.17\times 10^{-10}$ & $< 10^{-21}$ & $1.79\times 10^{-14}$ & $< 10^{-21}$ & $< 10^{-21}$ \\ 
            \midrule
            (255,13,59)	& 10 & $7.97\times 10^{-7}$ & $1.45\times 10^{-9}$ & $< 10^{-21}$  & $3.56\times 10^{-8}$ & $1.59\times 10^{-11}$ & $< 10^{-21}$ & $< 10^{-21}$ & $< 10^{-21}$ & $< 10^{-21}$ \\ 
			\midrule
			\bottomrule 
	\end{tabular}
    }
	\label{tab:Pfail}
	\begin{tablenotes}
		\item{The FRAM and SRAM memory can be reused when multiple BCH blocks and hash are sequentially computed.
        }
	\end{tablenotes}
\end{table*}

\begin{table*}[h]
	\centering 
	\caption{Overhead of RFE and FE when SRR, 2MRR and 3MRR are used.}
	\resizebox{1.0\textwidth}{!}{
		\begin{tabular}{c | c | c | c | c | || c | c | c || c | c }
			\toprule 
			\toprule 
			& & \multicolumn{3}{c|||}{Reverse Fuzzy Extractor} & \multicolumn{5}{c}{Fuzzy Extractor} \\
			\cmidrule(l){3-5} 
            \cmidrule(l){6-10}
            
			& & \multicolumn{1}{c}{CPU cycles}& \multicolumn{2}{c|||}{Memory Usage} & \multicolumn{3}{c}{CPU cycles} & \multicolumn{2}{c}{Memory Usage} \\
            \cmidrule(l){4-5}
			\cmidrule(l){6-8}
            \cmidrule(l){9-10}
			($ n_1 $,$ k_1 $,$ t_1 $) & \rotatebox{0}{block num.} &  & FRAM & SRAM & SRR & 2MRR & 3MRR & FRAM & SRAM \\ 
            \midrule
			(63,18,10)	& 8 & 745,721 & 5,819 bytes & 352 bytes & 3,360,134 & 6,720,268 & 10,080,402 & 6,843 bytes& 1,464 bytes \\ 
			\midrule
			(63,16,11)	& 8 & 722,193 & 5,699 bytes & 355 bytes & 3,689,662 & 7,379,324 & 11,068,986 & 6,983 bytes & 1,406 bytes \\ 
			\midrule
			(127,29,21)	& 5 & 1,221,359 & 6,019 bytes & 470 bytes & 8,081,576 & 16,163,152 & 24,244,728 & 11,563 bytes & 1,466 bytes\\ 
			\midrule
			(127,22,23)	& 6 & 1,319,223 & 6,033 bytes & 477 bytes & 10,663,102 & 21,326,204 & 31,989,306 & 12,095 bytes & 1,464 bytes\\ 
			\midrule
			(127,15,27)	& 9 & 1,316,184 & 6,015 bytes & 484 bytes & 19,129,444 & 38,258,888 & 57,388,332 & 13,159 bytes & 1,466 bytes \\ 
			\midrule
			(255,47,42)	& 3 & 2,063,241 & 6,407 bytes & 708 bytes & 18,515,476 & 37,030,952 & 55,546,428 & 28,925 bytes & 1,466 bytes \\ 
			\midrule
			(255,29,47)	& 5 & 2,200,269 & 6,467 bytes & 728 bytes & 35,091,151 & 70,182,302 & 105,273,453 & 31,535 bytes & 1,466 bytes \\ 
			\midrule
			(255,21,55)	& 7 & 2,377,650 & 6,481 bytes & 734 bytes & 58,631,390 & 117,262,780 & 175,894,170 & 31,535 bytes & 1,466 bytes \\ 
			\midrule
            (255,13,59)	& 10 & 2,329,519 & 6,379 bytes & 742 bytes & 85,493,076 & 170,986,152 & 256,479,228 & 39,527 bytes & 1,536 bytes \\ 
			\midrule
			\bottomrule 
	\end{tabular}
    }
	\label{tab:FEKeyGen}
	\begin{tablenotes}
		\item{The FRAM and SRAM memory can be reused when multiple BCH blocks and hash are sequentially computed.}
	\end{tablenotes}
\end{table*}
\noindent\textbf{Overhead}
We are now able to compare the overhead of MR$^3$FE (RFE with MRR) with the conventional RFE (only using a single reference response) when they are implemented in on a CRFID token. Considering performance advantages, BLAKE2s-128 is chosen for the hash function (cf. Table.~\ref{tab:hash} in the Appendix). Notably, RFE based mutual authentication requires a hash operation three times as highlighted in $\text{\ding{174}}$, $\text{\ding{175}}$ and $\text{\ding{176}}$ (cf. Fig.~\ref{fig:MRRFE}). In Table.~\ref{tab:FEKeyGen}, the overhead of RFE based mutual authentication is detailed when SRR, 2MRR, 3MRR are deployed. We observe the following:
\begin{itemize}
\item \noindent{\bf Single Readout Response Enrollment:} To achieve $\mathbb{P}_{\rm fail} < 10^{-6}$, ten BCH(255,13,59) blocks are required when the conventional single reference response under $25\celsius$ is used, whereas nine smaller BCH(127,15,27) blocks are adequate when 3MRR under $-15\celsius$, $25\celsius$, $80\celsius$ are deployed. In this context, the MR$^3$FE with 3MRR reduces clock cycle overhead by 43.50\% in comparison with a conventional RFE. 
\item  \noindent{\bf Majority Voting Response Enrollment:} To achieve $\mathbb{P}_{\rm fail} < 10^{-6}$, seven BCH(255,21,55) blocks are needed when the conventional single reference response under $25\celsius$ is used. In contrast, six smaller BCH(127,22,23) blocks are adequate when 3MRR under $-15\celsius$, $25\celsius$, $80\celsius$ are used. In this context, the MR$^3$FE with 3MRR reduces clock cycle overhead by 44.52\% in comparison with a conventional RFE. 
\item \noindent{\bf Pre-selection Response Enrollment:} To achieve $\mathbb{P}_{\rm fail} < 10^{-6}$, nine BCH(127,15,27) blocks must be applied when a conventional single reference response under $25\celsius$ is used. However, 8 smaller BCH(63,16,11) blocks are adequate when 3MRR under $-15\celsius$, $25\celsius$, $80\celsius$ are used. In this context, the MR$^3$FE with 3MRR reduces clock cycle overhead by 45.13\% in comparison with a conventional RFE.
\end{itemize}

Overall, we can summarize that the MRR always outperforms the SRR in the RFE case. Given the same overhead, the key failure rate is exponentially reduced via MRR. Alternatively, we can see that MRR greatly reduces the clock cycle overhead (nearly over 43\% reduction) to achieve same key failure rate as the conventional SRR method. Although memory usage reductions are not significant---memory is sequentially reusable, the ability of MRR to use smaller BCH codes lead to smaller code size and less internal state at run time.

\section{Multiple Reference Response Based Fuzzy Extractor (MR$^2$FE)}~\label{sec:MR2FE}
This section examines the generalization of the developed MRR methodology. We investigate the practicality of MRR when it is adopted to a FE scenario. In this context, the target device for implementation is not necessarily a highly resource constraint IoT device like the CRFID transponder. {\it We assume that the employment of a FE is mainly to derive a secure cryptographic key, while minimizing the FE implementation overhead is a desirable goal when possible}. In the case of MRR enabled FE---termed MR$^2$FE, it is the PUF device that iteratively carries out decoding and checking to identify whether the key is correctly recovered.

\subsection{MR$^2$FE}\label{Sec:M2RFE}
The MR$^2$FE key generator operates as follows:

\begin{enumerate}
\item During the key enrollment phase, the server registers $\{ {\bf r}_1$,...,${\bf r}_j$,..., ${\bf r}_J \}$, which are responses subject to the same challenge but generated under differing operating conditions: \{OC$_1$,...,OC$_j$,...,OC$_J$\}. Response ${\bf r}_j$ is hashed to gain a cryptographic key ${\bf sk}_j \leftarrow$\textsf{Hash}(${\bf r}_j$). The server computes helper data ${\bf p}_j \leftarrow${\textsf{Gen}(${\bf r}_j$)}, where $j\in \{1,...,J\}$.
\item During the key reconstruction phase, the PUF device regenerates response ${\bf r}^{\prime}$. Then the PUF device loads helper data ${\bf p}_j$, assumed to be public information and sent from the server at run time or loaded from an insecure off-chip/on-chip NVM. Simultaneously, the server sends a verification value $u_j \leftarrow$\textsf{Hash}(ID,$n_s$,${\bf sk}_j$,${\bf p}_j$) along with ID and $n_s$---the nonce generated by the TRNG in the server side----to the PUF device.
\item The PUF device performs ${\bf r}^{\prime}_j \leftarrow$\textsf{Rep}(${\bf p}_j$,${\bf r}^{\prime}$), where ${\bf r}^{\prime}_j$=${\bf r}_j$ only on the condition that $u^{\prime}_j$ =\textsf{Hash}(ID,$n_s$,${\bf sk}^{\prime}_j$,${\bf p}_j$)= $u_j$, with ${\bf sk}^{\prime}_j\leftarrow$\textsf{Hash}(${\bf r}^{\prime}_j$). Once this occurs, the PUF device is deemed to have successfully restored the response ${\bf r}_j$. Thus the secret key ${\bf sk}_j$ is recovered and {\it the following steps are skipped}.
\item Otherwise, if $u^{\prime}_j \neq u_j$, then the secret key ${\bf sk}_j$ reconstruction failed. Step 2) and 3) must be repeated for reconstructing another ${\bf sk}_i$ through ${\bf p}_i$, where $i\neq j$.
\item If none of the ${\bf sk}_j$, $j\forall \{1,...,J\}$, reconstructions is successful, then key reconstruction fails. 
\end{enumerate}

As we can see, the server actually enrolls $J$ helper data; each corresponding to one reference response generated under a varying operating condition. The PUF device iteratively conducts key reconstruction attempts to recover one enrolled cryptographic keys, ${\bf sk}_j$. If one of them is successfully reconstructed, the key recovery succeeds. Otherwise, if none of them succeeds, key reconstruction failure occurs.

To study the overhead of the MR$^2$FE key generator, we first comprehensively evaluate BCH code decoding overhead; this corresponds to the {\textsf{Rep}} implementation overhead.

\subsection{BCH Code Decoding}
BCH($ n_1 $, $ k_1 $, $ t_1 $) code is chosen again for consistency, its decoding overhead under different $ n_1 $, $ k_1 $, $ t_1 $ settings are tested, results are detailed in Table.~\ref{tab:BCH} in the Appendix. The experimental setup is the same as that described in Section~\ref{sec:TestSetup}.

\subsection{Comparison}
Following the operating steps of the MR$^2$FE key generator (cf. Section~\ref{Sec:M2RFE}), we are able to quantitatively compare the MR$^2$FE implementation overhead to token given SRR, 2MRR and 3MRR. Similar to Section~\ref{Sec:M3RFECompare}, we still select  BLAKE2s-128 hash function. MR$^2$FE needs to execute the \textsf{Rep} function {\it at most} $J$ times and the \textsf{Hash} function $2\times J$ times. This is significantly different from the MR$^3$FE where increasing $J$, the number of reference responses, brings {\it no extra} computational overhead to the token.

In practice, the secret key ${\bf sk}_j$ that is from a nominal operating condition, e.g., room temperature, is {\it advised} to be attempted first. Because, the ${\bf r}^\prime$ will be more likely reproduced under an operating condition that is close to the nominal operating condition. On the condition that such a trial succeeds, the remaining trials are no long needed---step 4) and 5) are  skipped, reducing overhead further. Nevertheless, we analyze the worst-case scenario, that is assuming that all $J$ trials have to be performed by the PUF device before a successful key recovery. Overhead comparisons for a FE with SRR, 2MRR, 3MRR are detailed in Table~\ref{tab:FEKeyGen}. We can make the following observations.
\vspace{2mm}

\noindent\textbf{Single Readout Response Enrollment:~} To achieve $\mathbb{P}_{\rm fail} < 10^{-6}$, ten BCH(255,13,59) blocks are required when the conventional single reference response under $25\celsius$ is used. In contrast, nine smaller BCH(127,15,27) blocks are adequate when 3MRR under $-15\celsius$, $25\celsius$, $80\celsius$ are used. In this context, the MR$^2$FE with 3MRR reduces clock overhead by 32.87\% in comparison with the conventional FE.
\vspace{2mm}

\noindent\textbf{Majority Voting Response Enrollment:~}
To achieve $\mathbb{P}_{\rm fail} < 10^{-6}$, seven BCH(255,21,55) blocks are required when the conventional single reference response under $25\celsius$ is used, whereas six smaller BCH(127,22,23) blocks are adequate when 3MRR under $-15\celsius$, $25\celsius$, $80\celsius$ are used. In this context, the MR$^2$FE with 3MRR reduces clock overhead by 45.44\% in comparison with conventional FE.
\vspace{2mm}

\noindent\textbf{Preselection Response Enrollment:~} To achieve $\mathbb{P}_{\rm fail} < 10^{-6}$, nine BCH(127,15,27) blocks are required when the conventional single reference response under $25\celsius$ is used. However, 8 smaller BCH(63,16,11) blocks are adequate when 3MRR under $-15\celsius$, $25\celsius$, $80\celsius$ are used. In this context, the MR$^2$FE with 3MRR reduces clock overhead by 42.14\% in comparison with a conventional FE.
\vspace{2mm}

MR$^2$FE still greatly outperforms the conventional FE with SRR. The reason lies on the fact that MR$^2$FE avoids the employment of a large BCH code block for decoding (BCH decoder complexity grows approximately with the square of the block size~\cite{yu2010secure}).  Therefore, given same key failure rate and a moderate number of model $J$, for example $J=3$, MR$^2$FE still consumes much less overhead even when it is the PUF device that has to perform $J$ trials. Overall, the overhead experimental results in Table~\ref{tab:FEKeyGen} demonstrates the efficacy gains from the MRR methodology also applies to a fuzzy extractor setting.

\section{Discussion}\label{sec:discussion}
In this section, first, we analyze security, including helper data manipulation attack, brute-force attacks and entropy leakage of (reverse) fuzzy extractors with MRR. Second, we show that MRR is not specific to the SRAM PUF, but appears to be generic to other silicon PUF types. Third, through hash and secure sketch overhead comparisons, we highlight the significance of reducing the overhead of secure sketch implementation for constructing a lightweight (reverse) fuzzy extractor. Then, we discuss limitations of our investigation and provide directions for future work.
\subsection{Helper Data Manipulation Attack}
Delvaux {\it et al.}~\cite{delvaux2014key,delvaux2014attacking}, first introduced helper data manipulation (HDM) attacks  although not on helper data generated from fuzzy extractors. In~\cite{delvaux2015helper}, HDM attacks are applied on a soft-decision error correction decoding. Here, the attacker sends manipulated helper data to the PUF key generator and observes key recovery failures. Over multiple queries, the attacker learns information about the PUF response, which eventually allows the attacker to recover the response. To prevent such an attack, one potential countermeasure is to check the integrity of helper data~\cite{delvaux2015helper}. During the key enrollment phase, the helper data $\bf p$ and the enrolled response $\bf r$ are hashed together to produce the hash value ${\bf u}$. The hash value $\mathbf{u}$ is validated during key recovery phase. Consequently, a helper data manipulation attack will always fail because the attacker is unable to provide a valid hash ${\bf u}$ since the attacker has no knowledge of the response $\bf r$.

Becker~\cite{becker2017robust} recently revealed a new HDM attack strategy against robust fuzzy extractors. In general, instead of attempting to recover the secret PUF response, the HDM attack attempts to set the PUF response corrected by the key generator to a response ${\bf r}_a$ predetermined by an attacker. Consequently, the attacker attempts to defeat the helper data integrity checks by crafting a hash value ${\bf u_a}$ and helper data ${\bf p_a}$ in an attempt to manipulate the PUF key generator with a high probability of producing the response ${\bf r_a}$ crafted by the attacker during the response error correction process. Such a HDM attack now allows an attacker to impersonate the PUF integrated device. Further, Becker's extended HDM attack allows the adversary to recover the original response $\bf r$, and the secret key. Various error correction codes including Reed-Muller codes~\cite{puchinger2015error,van2012soft} based on different decoding strategies, soft-decision codes~\cite{maes2009soft,van2012soft} and even-numbered repetition decodes~\cite{van2012soft} are examined and shown to be vulnerable~\cite{becker2017robust}.

A generic countermeasure against Becker's HDM attacks does not yet exist and remains an open challenge~\cite{becker2017robust}. However, the ability to mount the attacks depends on: i) the error-correction code employed; and ii) the method used for error-correction. Becker~\cite{becker2017robust} shows linear BCH code based syndrome decoding is immune to HDM attacks; an attacker is unable to set a specific response, although helper data may be manipulated to cause the corrected response bits to flip\footnote{ A detailed discussion and a proof that syndrome based decoding is immune from the HDM attacks presented by Becker can be found in Section 6.1 of the article in~\cite{becker2017robust}. Therein, Becker also derives a security criterion to validate the immunity of a decoding method against the HDM attacks presented in~\cite{becker2017robust}. }. Our evaluations on MR$^2$FE and MR$^3$FE is built upon BCH codes and employ syndrome decoding secure under HDM attacks. Nevertheless, the method we propose is~\textit{agnostic} to the fuzzy extractor employed, because we do not rely on any specific code or decoding strategy.  

Further, recall that helper data is manipulated and sent to the PUF device, while the attacker tests the key failure to determine their success. This implies that the attacker is able to conduct an arbitrary number of queries albeit less than the complexity of a brute force attack or random guessing. In a reverse fuzzy extractor setting, it is the PUF device or token performing the \textsf{Gen} operation to generate helper data. Thus, whenever a HDM attack is orchestrated by an attacker, the manipulated helper data is sent to the server. Consequently, HDM attacks target the server in the context of a reverse fuzzy extractor. In this setting, the HDM attack is likely to be detected by the resourceful server because of the abnormal key reconstruction failure rates resulting from the tampered helper data. We can observe that in a reverse fuzzy extractor setting, and when faced with a more intelligent server, an attacker will lose the assumed ability to apply an arbitrary number of queries.

\subsection{Entropy Leakage}
Given a secure sketch with BCH($n$,$k$,$t$) code, entropy leakage is caused mainly from the public helper data. The well-known min-entropy loss is the $n-k$ bound given the exposure of helper data. This $n-k$ bound is conservative. Research studies have explored the derivation of a tight bound for min-entropy loss~\cite{maes2016secure,delvaux2015helper,delvaux2016efficient}. However, calculation of min-entropy loss using the tight bound in~\cite{delvaux2016efficient} requires undertaking exhaustive simulations as a straightforward analytical equation is not available. The purpose of our work is to demonstrate that the MRR greatly reduces a token's (reverse) fuzzy extractor implementation overhead. Therefore, we consider the conservative ($n-k$) min-entropy loss bound where the public helper data generated by the BCH($n$,$k$,$t$) code leaks ($n-k$)-bit entropy~\cite{maes2016secure,delvaux2016efficient}. Then, taking response bias $b$ into account, the residual min entropy $\mathbb{H}_{\infty}$ of the $n$-bit response $\bf r$ conditioned on the public helper data $\bf p$ can be expressed as:
\begin{equation}\label{eq:entropyLeak}
\mathbb{H}_{\infty}(R|P)=n\cdot\log_2(max(b,1-b))-(n-k).
\end{equation}

The reverse fuzzy extractor and a conventional fuzzy extractor might not provide identical security guarantees expressed by Eq~\ref{eq:entropyLeak}. This is because a reverse fuzzy extractor can result in unanticipated entropy loss under repeated helper data exposure associated with a given PUF response $\bf r^{\prime}$; unless, PUF responses demonstrate a symmetry property. In other words, the one-probability, the probability of a given bit attaining a binary one value, of PUF responses is a symmetric distribution~\cite{kusters2017security}; alternatively, are unbiased. Generally, the extra entropy loss is a result of the leakage of bit-specific reliability information~\cite{delvaux2016efficient}.

The extra entropy loss is important only when PUF response bias is considerably different from the ideal value of 50\% as shown by the analysis in~\cite{delvaux2016efficient,delvaux2017physically}. We can see that for the SRAM PUFs tested in our work, the extra entropy loss is very small. For instance, at 25\degree~C the evaluated mean bias of SRAM PUFs we tested is 49.87\%. Such a small bias aligns with expectations from modern silicon PUFs according to other studies~\cite{roel2012physically}. In this context, employing a few more response bits in the reverse fuzzy extractor can compensate for the small extra loss in entropy. As observed by Delvaux~\cite{delvaux2017physically}, for a PUF with low bias within $[0.42, 0.58]$, increasing the length of raw responses alone is an effective measure.

If the bias is severe, entropy compensation by solely increasing the length of raw responses becomes ineffective. As a result, debiasing the biased raw responses~\cite{maes2016secure} must be undertaken first, e.g., via classic von Neumann (CVN) debiasing, pair-output von Neumann debiasing with erasures ($\epsilon$-2O-VN). Notably, not all debiasing schemes offer reusability---multiple use of the same PUF response---for a reverse fuzzy extractor~\cite{delvaux2016efficient}. Nevertheless, a reverse fuzzy extractor with MRR is naturally compatible with debiasing schemes that offer reusability.

\begin{figure}
	\centering
	\includegraphics[trim=0 0 0 0,clip,width=0.40\textwidth]{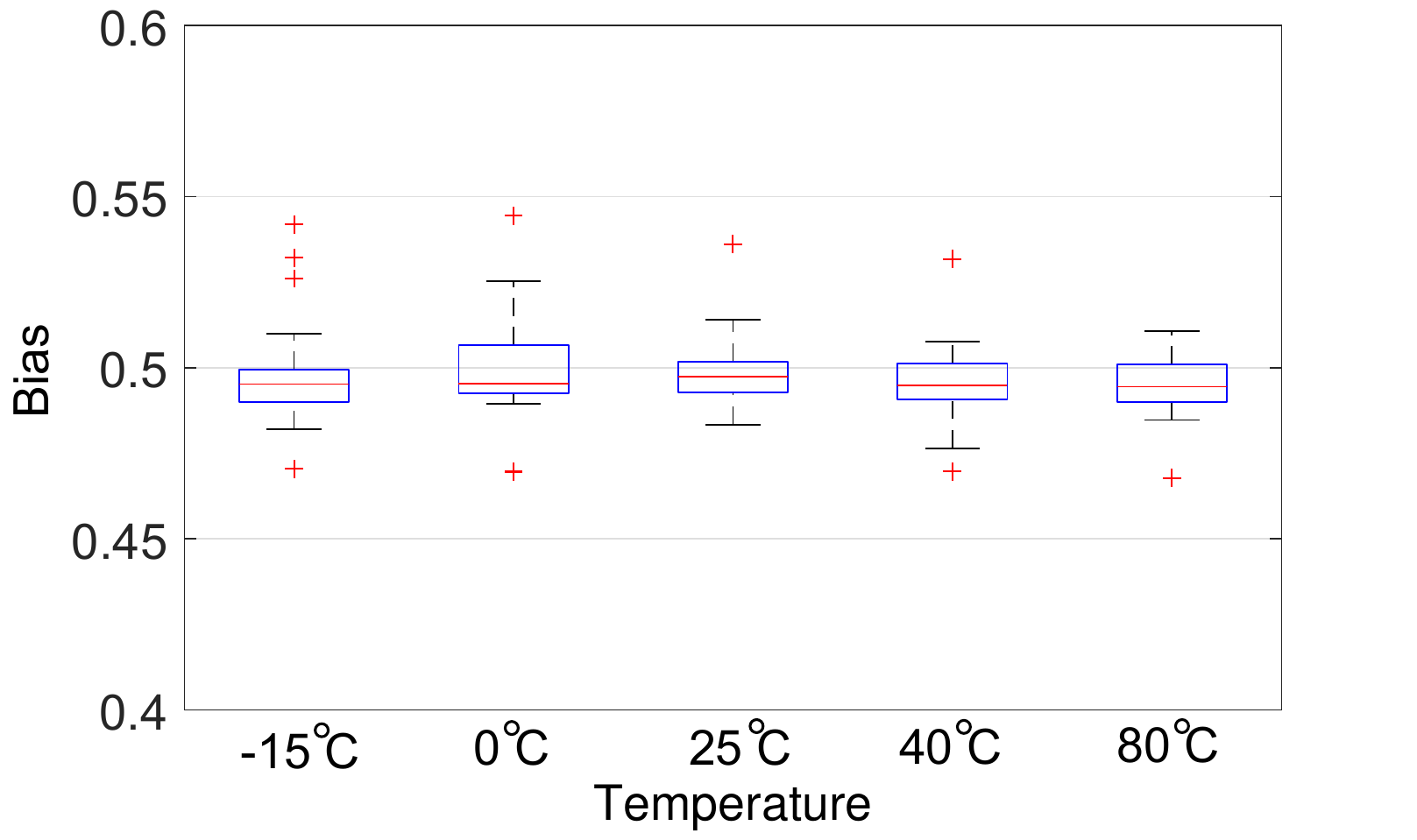}
	\caption{Bias of 23 tested SRAM PUFs under five different temperatures.}
	\label{fig:BiasDifTemp}
\end{figure}

We also tested the SRAM PUF bias in our dataset under the five different operating temperatures. We want to examine if there is a relationship between the bias---fraction of `1's---and the temperature. If there are more `1'/`0' responses, indicative of severe bias, when the temperature is higher or lower, then the temperature might cause an unanticipated entropy loss. The bias of 23 SRAM PUFs under varying temperature is detailed in Fig.~\ref{fig:BiasDifTemp}. We can observe that the mean bias is almost invariant to temperature. Therefore, we can expect that a change in the operating temperature to not lead to additional entropy loss.
 
\subsection{Brute Force Attack Complexity under MRR}
 We recognize that under a MRR model, whilst retaining the security properties of a given secure sketch, brute force attack complexity will reduce. Consider, using six parallel BCH(127,22,23) codes as outlined in Table.~\ref{tab:Pfail} and Table.~\ref{tab:FEKeyGen}. According to Eq.~\ref{eq:entropyLeak}, we can obtain a key ${\bf sk}$ having a 129-bit entropy given a bias of 0.4987 obtained from SRAM PUF test data. Therefore, without knowledge of $ {\bf sk} $, the probability of an adversary succeeding in a brute-force attack to determine $\textbf{sk}$ is $\frac{1}{2^{129}} $ when a conventional single reference response is employed in the reconstruction. When $J$ multiple response references are employed, the server or token can attempt to reconstruct the response, effectively, $J$ times to obtain the key $\bf sk$. Correspondingly, the brute-force attack complexity decreases linearly as a function of the number of multiple reference response models $J$; in our example, the probability of a brute force attack succeeding is $\frac{J}{2^{129}} $. In general, for $J$ multiple reference responses and min-entropy bound as expressed in Eq.~\ref{eq:entropyLeak}, we can express the probability $\mathbb{P}_{\rm{brute}}$  of a successful brute force attack as
\begin{equation}
\mathbb{P}_{\rm{brute}}=2^{-\mathbb{H}_{\infty}(R|P)+log_2(J)}. 
\end{equation}

Now, given $J$ MRR models, brute-force attack complexity reduces only slightly while significantly alleviating failure probability $\mathbb{P}_{\rm fail}$ and the token's \textsf{Gen} or \textsf{Rep} implementation overhead as shown in Table.~\ref{tab:Pfail} and Table.~\ref{tab:FEKeyGen}, respectively. For example, consider the probability of key failure $P_{\rm fail}=2.08\times10^{-5}$ for SRR and $5.47\times10^{-13}$ for 3MRR for the BCH (127,22,23) code in our example given in Table.~\ref{tab:Pfail}, when preselected response enrollment approached is utilized. Compared to the 3 fold increase in the success of a brute force attack, still extremely low, 3MRR has resulted in more than $10^7$ fold decrease in the key failure rate. Further, consider achieving a key failure rate of less than $10^{-6}$ using SRR. We can see from Table.~\ref{tab:Pfail} that 10 parallel blocks of BCH(255,13,59) have to be used when a single readout response enrollment is adopted. Considering the implementation overhead of a reverse fuzzy extractor on a token to also achieve key failure rate less than $10^{-6}$ employing MRR, nine smaller BCH(127,15,27) are adequate. In this context, the overhead in terms of CPU clock cycles is reduced   by 43\% with 3MRR---2,329,519 vs 1,316,184 clock cycles. We can see that the gains in failure rate and implementation overhead more than compensate for the very small reduction in brute force attack complexity.

\subsection{Generalizabilty of MRR}\label{sec:MRRGeneral}
We have validated the MRR method with SRAM PUFs where operating conditions that deviate from the reference condition results in responses that exhibit a higher BER than that responses generated at the enrolled reference condition. We can observe that this behavior agree with other  experimentally validated silicon PUFs such as ROPUFs~\cite{gunlu2015reliable}, Latch PUFs, D flip-flop PUFs, Buskeeper PUF and Arbiter PUFs~\cite{roel2012physically} summarized in Table. III and obtained from published literature. We can see that, in general, the BER increases when the difference between the reference operating condition and the condition under which the response is regenerated increases. We have shown in our study that in this context, our MRR approach provides flexibilities of selecting a reference operating condition that is potentially closer to the specific working operating condition of the in-the-field device. Therefore, in general, we can expect the MRR approach to provide an implementation efficiency for silicon PUFs.
\begin{table}[h!]
	\centering 
	\caption{BER evaluated as the difference between the enrolled and working operating conditions. The data is obtained from~\cite{roel2012physically}.} 
	\resizebox{0.5\textwidth}{!}{
	\begin{tabular}{c| c | c | c | c | c} 
					
		& \rotatebox{45}{({$-40\celsius$, $1.02$~V})} & \rotatebox{45}{({$-40\celsius$, $1.32$~V})} & \rotatebox{45}{ \begin{tabular}{@{}c@{}}Reference condition \\ ({$25\celsius$, $1.20$~V})\end{tabular} }& \rotatebox{45}{({$85\celsius$, $1.02$~V})}  & \rotatebox{45}{({$85\celsius$, $1.32$~V})}  \\ 
		\midrule
        Latch PUF & 23.10\% & 23.38\% & 2.61\% & 10.62\% & 10.60\% \\
        D Flip-flop PUF & 12.79\% & 12.90\% & 3.54\% & 18.10\% & 17.89\% \\
        Buskeeper PUF & 9.68\%\% & 9.77\% & 4.16\% & 17.71\% & 17.48\% \\
        Arbiter PUF & 7.41\% & 5.41\% & 3.04\% & 5.23\% & 5.34 \\
       RO PUF & 9.01\% & 7.81\% & 1.53\% & 7.11\% & 8.35\% \\
		\bottomrule
	\end{tabular}}
	\label{tab:BERLinear} 
\end{table}

\subsection{MRR Enrollment}
We observe that an improvement in the error correction efficiency is always achieved with trade-offs; for instance, increasing the enrollment overhead while reducing the key failure rate. We can see in a soft-decision decoding approach as in~\cite{maes2009low}, the PUF-key generator efficiency is enhanced but requires repeated response measurements, in the order of 10 to 100, to collect individual bit's reliability information as additional helper data~\cite{maes2009low}. A preprocessing method such as majority voting that can be used with hard decision decoding to reduce key failure rates also requires repeated measurements during the response bit enrollment phase~\cite{bohm2013puf}.

In our MRR approach, we trade-off the overhead of enrolling multiple reference responses during enrollment with a significantly reduced implementation efficiency on a token. We can also see that the corresponding implementation efficiency on a token does increase the computation burden on a server because multiple reference responses now need to be evaluated in parallel. More specifically, we can see that our MRR approach requires a moderately increasing enrollment overhead---enrolling $J$ reference responses given $J$ operating conditions \footnote{In practice, one can set up several temperature zones in one temperature oven---similar to the reflow oven. In this context, multiple reference responses are enrolled sequentially by passing the same PUF integrated IC into each temperature zone sequentially e.g., via a conveyor belt.} while it significantly reduces the encoding/decoding implementation overhead on a PUF token.

Our tested results demonstrate a small number of reference models, for example $J=3$ in our evaluation, already greatly improves (reverse) fuzzy extractor implementation efficiency on a token. For instance, to achieve a key failure rate $\mathbb{P}_{\rm fail} < 10^{-6}$, when a single readout response approach is utilized, the reverse fuzzy extractor with 3MRR reduces the clock cycle overhead with that of a conventional single reference response FE by 44.5\%. In addition, the MRR enrollment overhead is only incurred once during the enrollment phase but the benefits extends to the life of the token. We can see that MRR facilitates minimizing the token overhead. In the context of resource limited devices such IoT devices, or the CRFID token we have employed in our approach, minimizing implementation overhead is highly desirable in practice. We can expect a resourceful server in such a context to easily manage the increase in computation overhead.

\subsection{Hash and Encoding/Decoding Overhead}
Based on Table~\ref{tab:hash}, Table~\ref{tab:BCH} and Table~\ref{tab:BCHDEC}, we can see that it is paramount to minimize the absolute error correction overhead for not only the \textsf{Gen} but also \textsf{Rep} function because the overhead of a hash function is always much less than that of secure sketch. Or in other words, regardless of RFE or FE, the \textsf{Gen} and \textsf{Rec} implementation overhead is always dominant. According to the fully implemented PUF-based key generator on an FPGA platform by Maes {\it et al.}~\cite{maes2012pufky}, where concatenated (7, 1, 3) repetition code and a BCH(318, 174, 17) code was used, the BCH(318, 174, 17) decoder and (7, 1, 3) repetition code cost were 112 and 37 FPGA slices respectively, while the hash implementation of SPONGENT-128 only occupied up 22 slices. The hash function logic overhead was only 15\% of the error correction logic.

Our code encoding/decoding overhead evaluation agrees with this observation but from a different implementation. In particular, our results are from implementations in software rather than hardware~\cite{maes2012pufky}. In fact, in the software implementation of the hash and BCH decoder, the hash takes even less overhead in comparison with the BCH decoder. We can conclude that a lightweight PUF key generator is very hard to achieve without optimizing the error correction coding/decoding overhead regardless of implementation in hardware or software. The MRR method we present provides a new approach to substantially minimize the error correction overhead. 

\subsection{Limitations and Future Work}
\noindent\textbf{Generalizability:~}
In our experimental study, we have evaluated our MRR approach using SRAM PUFs. Although we discuss the generalizability to other silicon based PUF types in Section~\ref{sec:MRRGeneral} where the expected bit error rate  increases outside of the reference operating conditions, the MRR approach may not provide an implementation efficiency for PUF types that do not exhibit the above behavior. We believe the investigation of MRR to all PUF types raises an interesting research question to be addressed in the future.
\vspace{2mm}

\noindent\textbf{Soft-decision Decoding:~}
We have focused on hard-decision decoding and not considered the impact of our approach on soft-decision decoding. In general, we believe the MRR approach used with a soft-decision decoding strategy can further reduce implementation overhead. Consider, for example, the two approaches: i) soft-in-soft-out~\cite{maes2009soft,hiller2016cherry}; and ii) hard-in-soft-out~\cite{van2012soft} where information of bit-specific reliability is employed as helper data to point out reliable responses to improve the gain of the PUF key generator. Now, instead of enrolling the bit-specific reliability along with the response itself under a single reference operating condition, bit-specific reliability along with response itself under multiple reference operating conditions can be deployed. We can see that the MRR approach has the potential to help lower key the failure rate further and, consequently, reduce the footprint of the needed key generator implementation.

The extent to which the MRR approach can provide a benefit under soft-decision decoding should be investigated further. However, we can see that soft-in-soft-out decision decoding in~\cite{maes2009soft} and decoding strategy based on hard-in-soft-out proposed in~\cite{van2012soft} are found to be vulnerable to the helper data manipulation attacks proposed in~\cite{becker2017robust}. Therefore, care should be taken in selecting the soft-decision decoding methods for evaluation.
\vspace{2mm}

\noindent\textbf{A Tight Bound for Key Failure Rate with MRR:~}
 In our current work, we limit our evaluation of ${\mathbb{P}_{\rm fail}}$ to employing $min\{\mathbb{P}_{2j}\},~j\in\{1,...,J\}$ to demonstrate the MRR's efficacy in a very conservative setting. Tight bounds on evaluating ${\mathbb{P}_{\rm fail}}$ that considers the complexity of correlation between the reference responses leaves for very interesting future work. 
\begin{equation}\label{eq:Pfail_tight}
{\mathbb{P}_{\rm fail}}=Pr({\bf r}_1,\cap...\cap,{\bf r}_j,\cap...\cap,{\bf r}_J) ,~j\in\{1,...,J\}.
\end{equation}
\section{Conclusion}~\label{sec:conclusion}
We constructively developed MRR to significantly reduce the overhead of RFE and FE implementations and proposed  MR$^3$FE and MR$^2$FE for lightweight mutual authentication and key generation. We comprehensively validated our approach through a case study by implementing on an ultra low power MCU employed by a CRFID transponder (WISP5.1LGR) as an exemplary resource constrained IoT device. Our extensive experimental analysis demonstrate that, regardless of response enrollment approaches, (R)FE with MRR will always greatly outperforms the conventional (R)FE with a single reference response. Enrolling more reference responses under fine-grained operating conditions can further reduce a token's overhead, specifically, in MR$^3$FE case because its overhead is independent of the number of enrolled MRR. The proposed MRR is not only applicable to the case-studied SRAM PUF but also to other PUF types. Dedicated and specific implementation optimization, e.g, C code optimizations, can be exploited to further decrease the overhead we have reported in the paper.
\appendix
\begin{table}[h]
	\centering 
	\caption{Hash Overhead (message size$=240$ bytes).}
	\resizebox{0.5\textwidth}{!}{
	\begin{tabular}{c|| c || c || c || c} %
		\toprule 
		\toprule 
		
		name & digest size & clock cycles & FRAM usage & SRAM usage \\ 
		\midrule
	DM-SPECK64 & 64 bits & 178,448  & 1,566 bytes & 52 bytes  \\
	BMW-256 & 256 bits & 150,046  & 11,398 bytes & 215 bytes \\
	SHA1 & 160 bits & 159,969  & 12,442 bytes & 94 bytes \\
	BLAKE2s-256 & 256 bits & 106,482  & 4,964 bytes & 238 bytes  \\
	BLAKE2s-128 & 128 bits & 104,723  & 4,961 bytes & 238 bytes  \\
	SHA3-256 & 256 bits & 584,126  & 3,652 bytes & 472 bytes \\				
		\bottomrule
	\end{tabular}
    }
	\label{tab:hash} 
\end{table}

\begin{table}[h]
	\centering 
	\caption{BCH code encoding overhead.}
		\begin{tabular}{c|| c || c || c} %
			\toprule 
			\toprule 
			
			($ n_1 $,$ k_1 $,$ t_1 $) &  clock cycles & FRAM usage & SRAM usage \\ 
			\midrule
			(63,18,10) & 53,944  & 858 bytes & 114 bytes  \\
			(63,16,11) & 51,003  & 738 bytes & 117 bytes  \\
			(63,7,15) & 31,577  & 842 bytes & 126 bytes  \\ \midrule
			(127,64,10) & 238,671 & 858 bytes & 197 bytes  \\ 
			(127,57,11) & 248,433  & 1,002 bytes & 204 bytes  \\
			(127,50,13) & 235,005  & 1,034 bytes & 211 bytes  \\
			(127,43,14) & 219,095  & 1,050 bytes & 218 bytes  \\
			(127,36,15) & 198,758  & 1,044 bytes & 225 bytes  \\
			(127,29,21) & 181,438  & 1,058 bytes & 232 bytes  \\
			(127,22,23) & 167,509  & 1,072 bytes & 239 bytes  \\
			(127,15,27) & 111,335  & 1,054 bytes & 246 bytes  \\ \midrule
			(255,123,19) & 930,093  & 1,370 bytes & 394 bytes  \\
			(255,63,30) & 680,087  & 1,418 bytes & 454 bytes  \\
			(255,47,42) & 583,024  & 1,446 bytes & 470 bytes  \\
			(255,37,45) & 476,744  & 1,492 bytes & 480 bytes  \\
			(255,29,47) & 377,220  & 1,506 bytes & 488 bytes \\
			(255,21,55) & 294,783  & 1,520 bytes & 496 bytes  \\
            (255,13,59) & 201,535 & 1,418 bytes & 504 bytes \\
			\bottomrule
		\end{tabular}
		\label{tab:BCH} 
	\end{table}
    
    \begin{table}[h]
	\centering 
	\caption{BCH code decoding overhead.}
		\begin{tabular}{c|| c || c || c} %
			\toprule 
			\toprule 
			
			($ n_1 $,$ k_1 $,$ t_1 $) &  clock cycles & FRAM usage & SRAM usage \\ 
			\midrule
			(63,18,10) & 393,836  & 1,882 bytes & 1,226 bytes  \\
			(63,16,11) & 435,027  & 2,022 bytes & 1,168 bytes  \\
			(63,7,15) & 626,881  & 2,572 bytes & 1,154 bytes  \\ \midrule
			(127,64,10) & 670,622 & 3,676 bytes & 1,226 bytes  \\ 
			(127,57,11) & 748,933  & 3,942 bytes & 1,228 bytes  \\
			(127,50,13) & 1,030,444  & 4,466 bytes & 1,228 bytes  \\
			(127,43,14) & 978,775  & 4,728 bytes & 1,228 bytes  \\
			(127,36,15) & 1,057,415  & 5,006 bytes & 1,218 bytes  \\
			(127,29,21) & 1,574,426  & 6,602 bytes & 1,228 bytes  \\
			(127,22,23) & 1,742,276  & 7,134 bytes & 1,226 bytes  \\
			(127,15,27) & 2,102,222  & 8,198 bytes & 1,228 bytes  \\ \midrule
			(255,123,19) & 2,515,163  & 11,958 bytes & 1,228 bytes  \\
			(255,63,30) & 4,116,796  & 17,700 bytes & 1,228 bytes  \\
			(255,47,42) & 6,102,010  & 23,964 bytes & 1,228 bytes  \\
			(255,37,45) & 6,582,507  & 25,530 bytes & 1,226 bytes  \\
			(255,29,47) & 6,976,341  & 26,574 bytes & 1,228 bytes \\
			(255,21,55) & 8,345,992  & 30,750 bytes & 1,226 bytes  \\
            (255,13,59) & 8,528,363 & 34,566 bytes & 1,298 bytes\\
			\bottomrule
		\end{tabular}
		\label{tab:BCHDEC} 
	\end{table}

\end{document}